\documentclass[
aps,
final,
notitlepage,
oneside,
onecolumn,
nobibnotes,
nofootinbib,
superscriptaddress,
showpacs]
{revtex4}
\usepackage{graphicx}
\usepackage{amsmath}
\usepackage{amsfonts}
\usepackage{amssymb}
\usepackage{bbm}
\usepackage{yfonts}

\newcommand{\PL}{\protect{Pl{\"u}cker }}

\begin{document}

\title{On a Microscopic Representation of Space-Time III}
\author{\firstname{Rolf}~\surname{Dahm}}
\affiliation{Permanent address: beratung f{\"{u}}r IS, G{\"{a}}rtnergasse 1, D-55116 Mainz, Germany}
\email{dahm@bf-is.de}

\thanks{We thank J. G. Vargas for several interesting discussions
during the ICCA-10 conference and especially for pointing out various
aspects of K{\"a}hler's work as well as K{\"a}hler's achievements 
in differential calculus.}


\date{July 31, 2018}

\begin{abstract}
Using the Dirac (Clifford) algebra $\gamma^{\mu}$ as initial 
stage of our discussion, we summarize previous work with 
respect to the isomorphic 15dimensional Lie algebra su*(4) 
as complex embedding of sl(2,$\mathbb{H}$), the relation to 
the compact group SU(4) as well as subgroups and group chains.
The main subject, however, is to relate these technical 
procedures to the geometrical (and physical) background which 
we see in projective and especially in line geometry of $\mathbb{R}^{3}$.
This line geometrical description, however, leads to applications
and identifications of line Complexe and the discussion 
of technicalities versus identifications of classical line 
geometrical concepts, Dirac's 'square root of $p^{2}$', 
the discussion of dynamics and the association of physical 
concepts like electromagnetism and relativity. We outline 
a generalizable framework and concept, and we close with 
a short summary and outlook.
\end{abstract}

\maketitle
\section{Introduction}
\label{ch:intro}

\subsection{Context so far}
\label{sec:context}
In the first two parts (\cite{dahm:MRST1}, \cite{dahm:MRST2}) 
of this series of papers we've presented a mostly group-based 
approach to the Dirac algebra where we've started from nothing
but very basic assumptions of spin and isospin symmetries in 
order to describe hadronic observables in the low-energy regime
of the particle spectrum. The straightforward part of our approach
resulted in a compact SU(4) ($A_{3}$) group\footnote{In this 
energy regime, counting of (grouped) resonances works well with 
respect to dimensions of SU(4) group reps (references in 
\protect{\cite{dahm:MRST2}}).} covering independent SU(2)$\times$SU(2) 
spin$\times$isospin or isospin$\times$spin transformations, 
dependent on the respective operator representation (hereafter
for short 'rep') identifications.

As the main step, based on several observations, we've introduced 
{\it only one physical assumption}: We want to understand this 
compact SU(4) symmetry, although mathematically represented as
an {\it exact} symmetry, physically as a 'nonrelativistic' (or 'low 
energy') {\it approximative} limit of an appropriate relativistic
description in terms of SU$*$(4) $\cong$ Sl(2,$\mathbb{H}$), so 
we use compact SU(4) as a (physical) approximation or 'effective' 
description only in order to use its well-established rep theory
of compact Lie groups. With respect to the spectrum, we have to 
group 'particles' and 'resonances', so consequently we break the 
(noncompact and compact) symmetry of SU$*$(4) and SU(4), respectively,
further by spontaneous symmetry breaking with respect to the
Wigner-Weyl realized compact (maximal) subgroup USp(4) and 
other mechanisms later on.
So in \cite{dahm:MRST2}, we have continued this discussion (see 
also \cite{dahm:MRST1} and references) by presenting some more 
aspects with emphasis on spontaneously (and later explicitly) 
broken symmetries and some evidence to relate usual/standard 
quantum field theory to a background in projective and especially
line geometry. In general, by the Lie group/algebra considerations
in terms of symmetric spaces, so far we have obtained the three
reduction chains
\begin{equation}
\label{eq:chains}
\begin{array}{c}
\frac{\mathrm{SU}*\mathrm{(4)}}{\mathrm{USp(4)}}\times
\frac{\mathrm{USp(4)}}{\mathrm{SU(2)}\times\mathrm{U(1)}}\times
\mathrm{SU(2)}\times\mathrm{U(1)}\\[3mm]
\frac{\mathrm{SU}*\mathrm{(4)}}{\mathrm{USp(4)}}\times
\frac{\mathrm{USp(4)}}{\mathrm{U(2)}}\times
\mathrm{U(2)}\\[3mm]
\frac{\mathrm{Sl(2,}\mathbb{H}\mathrm{)}}{\mathrm{U(2,}\mathbb{H}\mathrm{)}}\times
\frac{\mathrm{U(2,}\mathbb{H}\mathrm{)}}{\mathrm{Gl(1,}\mathbb{H}\mathrm{)}}\times
\mathrm{Gl(1,}\mathbb{H}\mathrm{)}\\
\end{array}
\end{equation}
on complex and quaternionic spaces, respectively (\cite{dahm:MRST1}
and references). The physical interpretation, however, has to be
worked out step by step, and separately per chain. With respect to
the first 'quotients' of all three chains, as a first approach due
to the concept of spontaneous symmetry breaking and the occurrence 
of related 'Goldstone' bosons, we've focused on the 5-dim coset 
space given by
\begin{equation}
\label{eq:cosetrep}
\exp p\,=\,
\exp\left\{p_{A}\mathcal{P}_{A}\right\}\,=\,
\tilde{p}_{0}\mathbbm{1}+\tilde{p}_{A}\mathcal{P}_{A}\,,\,
1\leq A\leq 5\,,
\end{equation}
where $\mathcal{P}_{A}=\{iQ_{01},iQ_{03},Q_{12},Q_{22},Q_{32}\}$
in our usual twofold quaternionic basis (see \cite{dahm:MRST1} 
and references) and its SO(5,1) symmetry with respect to 
$(\tilde{p}_{0},\tilde{p}_{A})$ \cite{dahm:MRST1}. Whereas the
mathematical background is the double covering of SO(5,1) by
SU$*$(4) $\cong$ Sl(2,$\mathbb{H}$), the only physically feasible
identification has been the association of the photon, its soft
scattering limit and the occurrence of Bremsstrahlung when changing
velocities of charged particles. So we've interpreted the necessary
5-dim Goldstone boson(s) of SU$*$(4)/USp(4) not as usual in terms 
of five individual fields, but as a single, 5-dim line rep of 
a base element in line geometry (or $P^{5}$, respectively). 
Please note once more, that this discussion is {\it not} restricted
to the old (and sometimes simple) spin/isospin hadron interpretation
of the reps (see e.g.~\cite{bjoedrell}) but holds for {\it all} 
quantum theoretical descriptions based on the Dirac (Clifford) 
algebra due to its isomorphism\footnote{We have addressed the 
problem already that there are various compact low-dimensional
symmetry groups which occur automatically in this context. So 
there is a priori {\it no need} to introduce manually (and 
additionally) further degrees of freedom based on such groups
by hand like in gauge or Yang-Mills approaches but it is more
important to gain control over the respective (physical) field
interpretations \protect{\cite{dahm:QTS7}} to avoid superfluous
degrees of freedom and double counting. The decomposition 
(\ref{eq:chains}) {\it already contains} SU(2)$\times$U(1)
(or its covering Gl(1,$\mathbb{H}$), respectively), there is
a priori no need to introduce additional fields by adding 
additional SU(2)$\times$U(1) structure.} with SU$*$(4).

In this context, we've begun branching into a parallel thread
(see the conference contributions \cite{dahm:QTS7} and 
\cite{dahm:GOL}) which led deeper into projective geometry 
and transfer principles, and as such to various equivalent 
representations of geometries (see e.g. \cite{blaschkeIII:1928}).
In terms of (Lie) group theory, we are thus dealing not only
with SO(5,1), but with the real groups SO($n$,$m$) where $n+m=6$ 
and -- by complexifying some of the coordinates in use -- with
various (complex or quaternionic) covering groups and their 
subgroups. As such, we find on various levels correspondences
between group transformations and reps on one side as well as
geometries and objects on the other side.

\subsection{Outline}
At this stage of work, we want to present some more remarks
on physical aspects of a quaternionic projective theory
(QPT, see \cite{dahm:MRST1} and references therein), and we 
try to relate them to geometrical concepts. Although at a 
first glance this seems like rewriting some 'well-known' 
representations only, in the long run, we benefit from a 
well-defined and unique description in terms of line (and 
Complex\footnote{As before, we have used Pl{\"u}cker's old 
German notation 'Complex' \protect{\cite{plueckerNG:1868}}
with capital 'C' to denote line Complexe, and as such we have
also used the old German plural form 'Complexe'. So mix-ups 
with complex numbers are (hopefully) avoided, moreover it 
would be nice to honour this great scientist (although late)
by using and establishing at least this small part of his 
notation.}) coordinates and their more general justification
right from projective geometry by using 4-dim lines as basic
space elements of $\mathbb{R}^{3}$ or $P^{3}$ instead of 3-dim
points and planes. Note, that this slightly different approach
is based on Pl{\"u}cker's identification of using $n$-dim 
objects in general as geometrical base objects \cite{plueckerNG:1868}
in 3-dim space. In the background, by using lines and Complexe,
we work with $P^{5}$, however, restricted to $P^{3}$ (or one
of the geometries in $\mathbb{R}^{3}$ if we impose the \PL 
condition (see eq.~(\ref{eq:plueckercondition})) on the six
line coordinates which also restricts the elements of $P^{5}$ 
to the Pl{\"u}cker-Klein quadric in $P^{5}$. So a quadratic 
constraint in $P^{5}$ governs the reps in $P^{3}$, and it
influences the relevant algebras in $P^{3}$ as well.

Last not least, lines in the context of tangential and 
especially tetrahedral Complexe automatically (and naturally)
introduce harmonic ratios\footnote{German: Doppelverh{\"a}ltnisse}
of points (and as such naturally metric properties from the
viewpoint of Caley-Klein metrics), not to mention polar and 
conjugation relations and a discussion of second order/class 
properties. Although here we do not have room to discuss many 
details of our ongoing work, we want mention at least some
direct relations with respect to electrodynamics and relativity.

As such, in the subsections of this first section, we'll summarize 
briefly some basic concepts and notations which we need throughout
this presentation. Afterwards, in section~\ref{ch:physics}, we
switch to selected physical aspects and identifications in order
to attach some well-known physical concepts -- however, usually
represented in analytical point descriptions -- to this alternative
approach by line representations, and sets thereof. Note already
here, that describing 3-dim space by points only, and without
comprising planes, is incomplete in that one neglects correlations,
or 'duality', or the adjoint/transposed reps, respectively.
However, an approach by lines formally comprises a priori both
types of transformations, collineations {\it and} correlations,
as duality in 3-dim space connects lines to lines. The cost 
of this intrinsic formal 'completeness' is a more difficult 
physical interpretation of the objects and transformations,
and quadratic constraints, as we have to consider dual/conjugated
lines on the same footing, so one has to put special emphasis 
on treating involutions correctly. 

Section~\ref{ch:program} thus steps back from details to allow
for a general view on the framework of Complexe, i.e.~a line-based 
description of 3-dim space, as far as we understand those 
connections, and it tries to shape a program which we plan
to pursue in our upcoming papers and publications. Please remember
throughout this 'program' that we do not reinvent mathematics 
and geometry, but that we want to argue in favour of lines, 
Complexe and spheres instead of points and planes only, because
we feel lines and spheres (as well as their assemblies) much
more suited to describe physics and physical observations than
the 'classical' point picture.
In the last section, we close with a brief summary and outlook of 
ongoing work.

\subsection{Summary \PL and Line Coordinates}
In order to discuss physics in terms of line geometry, it is
helpful to recall some basic notations\footnote{A longer 
derivation of various line coordinates right from the underlying
coordinate projections, i.e.~starting in terms of inhomogeneous
coordinates, can be found in \protect{\cite{plueckerNG:1868}}.
Take care, however, of the orientation of the underlying coordinate
system.}. Using four real homogeneous {\it point} coordinates $x_{\alpha}$, 
$0\leq\alpha\leq 3$, to denote points in projective 3-space $P^{3}$, 
by choosing two points $x$ and $y$ incident with the line, we 
can define the six independent (homogeneous) \PL coordinates\footnote{To
denote line coordinates, we use Study's notation with capital 
fracture letters.} $\mathcal{X}_{\alpha\beta}$ of the line by 
\begin{equation}
\label{eq:plueckervars}
\mathcal{X}_{\alpha\beta}:=x_{\alpha}y_{\beta}-x_{\beta}y_{\alpha}
\quad\mathrm{or}\quad
\mathcal{X}_{\alpha\beta}:=
\left|
\begin{array}{cc}
x_{\alpha} & y_{\alpha}\\
x_{\beta} & y_{\beta}
\end{array}
\right|
\end{equation}
where $0\leq\alpha,\beta\leq 3$. The coordinates are antisymmetric,
i.e. $\mathcal{X}_{\beta\alpha}=-\mathcal{X}_{\alpha\beta}$, 
invariant under common (additive) displacement of both points,
and they fulfil the '\PL condition'
\begin{equation}
\label{eq:plueckercondition}
P\,=\,
\mathcal{X}_{01}\mathcal{X}_{23}+\mathcal{X}_{02}\mathcal{X}_{31}+\mathcal{X}_{03}\mathcal{X}_{12}=0\,.
\end{equation}
Moreover, they transform linearly and homogeneously with respect 
to projective (space) transformations $a_{\alpha\beta}$, i.e.
\begin{equation}
\mathcal{X}'_{\alpha\beta}=\sum a_{\alpha\mu}a_{\beta\nu}\mathcal{X}_{\mu\nu}\,,
\end{equation}
so that for line coordinates $\mathcal{X}_{\alpha\beta}$, we 
may use a '6-dim' 'linear' rep $p_{A}$, $0\leq A\leq 6$, with
special constraints as well.
Given two lines with coordinates $\mathcal{X}_{\alpha\beta}$ and 
$\mathcal{X}'_{\alpha\beta}$, the incidence relation of the two 
lines reads as (\cite{klein:1872a} \S 1, polar equation)
\begin{equation}
\label{eq:lineincidence}
\mathcal{X}_{12}\mathcal{X}'_{34}
+\mathcal{X}_{13}\mathcal{X}'_{42}
+\mathcal{X}_{14}\mathcal{X}'_{23}
+\mathcal{X}_{34}\mathcal{X}'_{12}
+\mathcal{X}_{42}\mathcal{X}'_{13}
+\mathcal{X}_{23}\mathcal{X}'_{14}=0\,.
\end{equation}
This can be obtained by differentiating the \PL condition $P=0$
in eq.~(\ref{eq:plueckercondition}) according to
\[
\sum\frac{\partial P}{\partial \mathcal{X}_{\alpha\beta}}\cdot
\mathcal{X}'_{\alpha\beta}\,=\,0\,.
\]

The second definition on the rhs of eq.~(\ref{eq:plueckervars}),
and in general the determinant (re-)formulation -- at that 
time being more of a fashion -- is easier to relate to 
symplectic transformations. By transfer principles\footnote{We
follow Klein \cite{klein:1872a} with respect to his exposition
related to the Pl{\"u}cker-Klein quadric, however, with respect
to his stereographic projection(s) and coordinate discussion,
we want to postpone the discussion. Note, that in this paragraph,
we've adopted his notation of $\mathbb{R}^{3}$ and $\mathbb{R}^{5}$
instead of differentiating further with respect to projective
coordinates and spaces.}, the line and Complex geometry of 
$\mathbb{R}^{3}$ can be mapped onto points in $\mathbb{R}^{5}$,
and we can perform analogous (and sometimes easier) point 
considerations in $\mathbb{R}^{5}$ where the Pl{\"u}cker-Klein 
quadric $M_{4}^{2}$ plays an important r\^{o}le \cite{klein:1872a} 
\cite{dahm:GOL} in that lines in $\mathbb{R}^3$ are points of
$\mathbb{R}^{5}$ located on the Pl{\"u}cker-Klein quadric 
$M_{4}^{2}$. Investigating images of objects and transformations
of $\mathbb{R}^3$ also in $\mathbb{R}^{5}$, special attention 
can be given to automorphisms of $M_{4}^{2}$ \cite{dahm:GOL}
\cite{study:1903}. The transition to other geometries (like 
Laguerre, M\"{o}bius, spheres, etc.) and related geometrical
objects \cite{blaschkeIII:1928} may be performed as well. Another 
closely related and deeply entangled aspect of line coordinates
is Pl{\"u}cker's notion of a Complex \cite{plueckerNG:1868}
(and M{\"o}bius' null systems in relation to planar lines of 
a linear Complex) as well as the related general geometry of 
Complexe and Congruences \cite{plueckerNG:1868}. 

In order to relate to (standard) differential geometry, it 
is easier to start right from Pl{\"u}cker's (Euclidean) 
coordinate rep (\cite{plueckerNG:1868}, p. 26, Nr. 26, eq.~(1))
\begin{equation}
\label{eq:linecoordpluecker}
(x-x'), (y-y'), (z-z'), (yz'-zy'), (zx'-xz'), (xy'-yx')
\end{equation}
of line (ray) coordinates\footnote{\PL usually used $(x,y,z)$ 
to denote coordinates of one single point $p$ instead of using
subscripts/indices attached to points $x$ or $y$ according to
$x_{i}$, $y_{i}$, etc.}.
If we now (in the sense of continuity and analyticity, or even 
associating a 'transformation' to 'connect' the two points
$x$ and $x'$ involved) require $x'_{i}=x_{i}+\mathrm{d}x_{i}$, 
i.e. $\mathrm{d}x_{i}=x'_{i}-x_{i}$, antisymmetry of the line 
coordinates with respect to point exchange $x'_{i}\longleftrightarrow x_{i}$
(or the equivalent description by a determinant when exchanging
columns) provides expressions in terms of coordinates and 
differential forms which directly lead to line elements 
$\overline{x'x}$, Pfaffian equations and the {\it calculus}\footnote{Note
the important fact that we need {\it a calculus} to reflect the
antisymmetry of the two points $x'_{i}$and $x_{i}$ involved,
and that in the context of $\mathrm{d}x_{i}$ we are talking 
of a calculus only!} of differential forms. 
If in addition, we introduce polar relations (i.e. we replace 
$\mathrm{d}x_{i}$ by brute force with the tangential 'operators' 
$\partial_{i}$ at the original -- and then only remaining 
and unique -- reference point $x$ of the tangent space, we
obtain (partial) differential representations of (compact)
Lie generators (see e.g.~\cite{gilmore:1974} or \cite{helgason:1978}
for their differential rep) according to $x_{i}\partial_{j}-x_{j}\partial_{i}$.
Up to coordinate complexifications which we'll discuss later,
the important fact, however, is the underlying geometry 
which is nothing but line geometry. We can use line geometry
to describe global/finite geometry, not only typical infinitesimal
problems and considerations, while maintaining full control
over the two points $x'_{i}$ and $x_{i}$ from above {\it individually}.
Working with finite points $x'_{i}$ and $x_{i}$, we may
treat also more advanced projective concepts like polarity
etc., and differential geometry can be considered as a
special concept only which may be derived always by 
well-defined limits and assumptions.

For us, it is noteworthy that the coordinate differences
$x'_{i}-x_{i}$ (see \ref{eq:linecoordpluecker} or \cite{dahm:GOL}),
on the one hand, show well-defined (line) transformation 
behaviour and a well-established geometrical interpretation
as projections, on the other hand, typical 'coordinate' 
transformations $\delta x_{i} = x'_{i}-x_{i}\sim \mathrm{d}x_{i}$
can be mapped to known Lie algebraic transformation concepts
like $\delta_{Y}\sim[Y,\cdot]$ or to transformations of 
differential forms. As such, also advanced algebraical 
and analytical concepts of such calculuses can be re-transferred
back to (projective) geometry\footnote{We thank J. G. Vargas
for pointing us to K\"{a}hler's work (see e.g.~\cite{kaehler:1960})
which we find really interesting to study in more detail 
also in the context of line geometry.}, and especially to 
line transformations and line geometry. So we think, line 
(and Complex) geometry is much better suited to describe 
physics than the various (infinitesimal and restricted) 
concepts 'derived' from differential geometry only. 

\subsection{'The Metric'}
\label{sec:themetric}
Above, we have mentioned already the mixture in notion 
nowadays when working with vectors as well as the
sometimes misleading (and most often 'vector-derived') 
notion and the intrinsic use of a metric. In most cases 
a 'vector', although formally a coordinate {\it difference},
is used by setting one of the two points to coincide with
the origin $0$ of a 'coordinate system'. This often shrinks
the coordinate difference to single point coordinates 
only, which afterwards often spoils the concept. The 
notion 'metric' -- whether in the usual Euclidean sense
or in the framework of (semi-)Riemannian spaces related
to differentials and 'line elements' -- usually describes
a symmetric (and diagonal) structure which is used to 
'contract' indices of two objects (e.g.~vectorial or 
tensorial reps) which themselves transform linearly. In 
most cases this notation is nowadays used in conjunction
with linear reps (on spaces/modules), and it is a fashion 
to discuss low-dimensional rep dimensions in the beginning 
and generalize soon to arbitrary (and sometimes infinite) 
rep dimensions. Typical examples are space-time using 
$x_{A}$, $A\geq 4$, and the related dynamics in various 
formulations, usually based on related 'momenta' $p_{A}$, 
$A\geq 4$, when applying Hamiltonian dynamics or 'quantum' 
approaches, and even 'time'-associated Lagrangean concepts
and (partial) differential equations. 

It is often overseen when starting from coordinates only 
and counting the coordinates naively, that already switching
the coordinate {\it interpretation} changes the 'dimension' 
of such objects or of the underlying rep space. Simple examples
are e.g. given by the 5-dim coset space $p$ (or $\exp p$)
in eq.~(\ref{eq:cosetrep}) when switching from 'space-time'
(point) interpretation as usual in nonlinear sigma models 
(or SSB models) to (infinitesimal) line elements (Lie), 
lines, Complexe or even more sophisticated geometrical 
models\footnote{See e.g.~\protect{\cite{enriques:1903}},
appendix II on 'abstract geometry' related to dim 5!}.

This eclipses the fact that in order to perform physics, 
we identify observable objects with special mathematical 
reps, and we map their (transformation) behaviour to reps 
having finite dimension only\footnote{i.e. the reps depend
on a finite number of parameters only!}. The same holds 
for well-established projective concepts like polarity 
and duality whose interpretations, when associated with 
physical objects, are often messed up by a generalization 
to arbitrary dimensions although one is -- at least 
sometimes -- still able to define a similar formal 
calculus and calculate 'for arbitrary dimension $n$'. 
It should be mentioned here that it is often the background 
of projective (or incidence) geometry within its application 
and usefulness for logic (see e.g. \cite{enriques:1903} with
respect to 'abstract geometry', there especially appendix II)
which still provides the helpful axiomatic background.

Here, we restrict our discussion to $\mathbb{R}^{3}$ 
where we have 3- or 4-dimensional (linear) point reps,
dependent on whether we use inhomogeneous or 
homogeneous/projective coordinates. By means of this
coordinate interpretation, already the line reps may
have 'dimension' 4, 5 or 6 \cite{plueckerNG:1868}, and
we know from the very beginning of projective geometry,
from duality, from the projective construction of objects
(e.g.~conic sections), or more general from synthetic 
geometry, that we can switch from using orders to using
classes, thus interrelating dimensions. So using $P^{3}$
as well as simple and well-known geometric objects, we 
are far from using only 3- or 4-dim reps to describe 
space-time objects and behaviour, and we find -- even 
on real spaces -- much more symmetry structure than 
simple transformation groups like SO(3) or the 
Poincar\'{e} group only \cite{dahm:GOL}. 

On this footing, interpreting as usual the (3-dim) momentum 
$\vec{p}$ as (polar part of a) line rep, and $\vec{x}^{\,2}=r^{2}=(ct)^{2}$ 
as a sphere with (infinite) radius per given common time 't' 
for all (projective!) space-dimensions $x_{i}$ and $x_{0}$, 
it is natural to (re-)introduce line coordinates as a unifying 
description which automatically comprises 'non-local' effects. 
The only price we have apparently to pay is a loss of the
direct (physical/metrical) coordinate interpretation of 
$x_{\alpha}$ like in Euclidean coordinates as well as a loss
of naive 1-dim parameter differentiation within the general 
geometric approach to describe point trajectories and/or orbits.
This 1-dim and mostly differential geometric aspect, however,
can be recovered by transitioning from (general) lines to line
elements while restricting the geometry and coordinatizing the 
respective geometrical setup by appropriately chosen (inhomogeneous)
coordinates. Moreover, this is just what Lie did when establishing
'Lie algebras' and the differential rep of generators\footnote{We 
think that this achievement is tightly related to having got 
knowledge on Pl{\"u}cker's work and establishing intensive contact
to Klein after having met Klein for the first time in October
1869 during Klein's 'Berlin time' from August 1869 to March
1870. However, we want to leave the (more) complete and final 
discussion and judgement to science historians.}. So we 
feel free to work with line geometry (or in some places 
even with the fully-fledged framework of projective geometry),
and we want to see how we can describe physics.

\subsection{Squares and Norms}
In this context, it is natural to understand the 'norms' 
$\vec{p}^{\,2}$ and $p^{2}=p_{\mu}p^{\mu}$ in terms of a 
{\it square} of (a part of) a line rep and as such -- 
remembering self-duality of lines in $\mathbb{R}^{3}$ 
-- when linearizing such squares we have to end up with 
a (linear) 5- or 6-dim line rep instead of a 3-dim 'vector' 
only\footnote{Thanks to his talk and private communication 
with O. Conradt during the conference (ICCA 10), we heard
that Dirac knew much about projective geometry, and that
it was Dirac who searched for (algebraic) reps of his 
results from within projective geometry. However, we do 
not have access to those references yet.}. So within the 
standard treatment, using the (3-dim) 'vector' approach
only, parts of the (6-dim) momentum rep (and as such 
moments and ('axial') parts of the energy) are often 
missing and are not considered in calculations. The same
holds for bilinear representations of a 'metric' in order
to linearize quadratic objects like in Clifford algebras 
or on semi-Riemannian spaces.

A naive generalization in terms of arbitrary dimensions 
spoils the background, i.e.~although formally we can 
rewrite (in Euclidean interpretation or using the 
four-vector calculus of special relativity) $p^{2}=m^2$
in terms of a linear 'vector' rep $p$ and a symmetric 
formalism $\{\gamma_{i},\gamma_{j}\}=\delta_{ij}$ or 
$\{\gamma_{\mu},\gamma_{\nu}\}=g_{\mu\nu}$ (see 
e.g.~\cite{bjoedrell} or \cite{lurie:1968}, ch.~1-3), 
the simple (formal) abstraction of a metric is algebraically 
nice to handle but too simple in order to highlight 
the complete geometrical (polar) background of such 
an 'anticommutator'. Of course, one finds an appropriate 
algebraic and analytic calculus like in Dirac's case,
and a generalization to arbitrary $n$ with lot of nice
algebra and group theory attached, but -- as history 
shows -- the fact that 6-dim line reps can be composed
of two 3-dim 'vectors' ('polar' and 'axial'), and as 
such exhibit naturally a SO(3)$\times$SO(3) transformation
structure, seems forgotten nowadays. Even worse, allowing
for individual coordinate complexifications (as long as 
we preserve the real 'norm' constraint $v_{i}^{2}=\mathrm{const}$)
for both of the 3-dim 'vectors' $v_{i}$ of a line rep, 
we can as well discuss SU(2)$\times$SU(2) or twofold 
quaternionic transformations U(1,$\mathbb{H}$)$\times$U(1,$\mathbb{H}$)
acting on these constituents, but we know the reason
for the different polar and axial behaviour of the
constituents by going back to eq.~(\ref{eq:linecoordpluecker}).
They result from emphasizing the absolute plane $x_{0}=0$
in projective geometry, thus defining affine and Euclidean
coordinates, i.e.~this decomposition in real 3-space is
an artefact of an effective description in terms of Euclidean
coordinates whereas the general theory to handle the description
thoroughly should be at least affine geometry, if not projective
geometry. So the discussions of chiral symmetry and chirality
fade out in front of this background of lines, linear Complexe,
and screws. Moreover, this raises the need for a thorough 
treatment of 'the metric' and especially of the geometrical
transition steps 'projective' $\longrightarrow$ 'affine'
$\longrightarrow$ 'Euclidean', and their analytical 
counterparts focusing on the changing coordinate 
interpretations and their respective analytical dependencies.

Last not least, this outlines our intention and motivation
to revive line and projective geometry instead of following
the usual 'linearization' of $p_{\mu}p^{\mu}=m^{2}$ by\footnote{We 
just want to remember the fact that this equation is
independent from the mass as $m$ drops out. Indeed, we
see this as an equation for 4-velocities $u_{\mu}$ 
describing the velocity constraint $u_{\mu}u^{\mu}=1$
(see also section~\ref{sec:metricNEW}, eq.~(\ref{eq:absolutequadric})).} 
$p_{\mu}\gamma^{\mu}$ discussing 'quantum' 'anything' 
and attaching algebra and analysis naively in form of 
one or the other calculus. For us -- arguing in Pl{\"u}cker's 
sense\footnote{It is a pity that the enormous achievements
of this great scientist are not only not honoured but 
even almost forgotten. To top this deficit, even his 
own university was able to publish only a short note 
\protect{\cite{bonnueberpluecker}} to remember his 140th 
anniversary of death in 2008. Even there, they put more 
focus on his CV and his 'strong and own' personality 
than on his enormous achievements in mathematics and 
physics (see e.g.~\protect{\cite{clebschpluecker}}). 
Indeed a lot of Pl{\"u}cker's results were absorbed 
later in Lie's, Klein's, Clifford's and Ball's work 
mentioning \PL only in general, or even without citing
or even mentioning \PL at all. This might be attributed
to the fact that \PL inbetween worked for decades in 
physics (and especially optics) only, before returning 
during the mid 1860s to mathematics while advising 
Klein in physics and mathematics. It was Klein in 
conjunction with Clebsch to summarize at least some 
of Pl{\"u}cker's late and more systematic results on 
line geometry \protect{\cite{plueckerNG:1868}}, based 
on existing manuscripts and on the outline originating 
from \PL, while \PL himself only had time to publish 
two late presentations on generalizations of lines to
'Complexe', 'Dynamen' and their tremendous use for 
physics before his death. For example, the treatment
of oval surfaces in relation to generating line sets
can be found in \cite{klein:1928} (see e.g.~ch.~II,
\S\S 4--6) or some very powerful consequences with 
respect to dynamics, differential geometry and cones
have been given by Clebsch in \protect{\cite{clebsch:1869}}\ldots}
-- the difference is the necessary switch towards 
using lines instead of only points (even if accompanied
by planes and duality) as the underlying base elements 
of space where people perform all kinds of analysis 
in 'space-time', even in terms of very sophisticated
concepts of differential geometry (see e.g.~\cite{percacci:1986})
which -- in our opinion -- hide more physics behind 
formal mathematics than they are able to show or describe.

\subsection{Summary 'Spheres' and Complexe}
\label{sec:plueckeroptics}
The most important aspect in our current context\footnote{
Please note, that the expression \protect{$p:=\vec{x}^{\,2}-r^{2}$}
is known as 'potency' (German: 'Potenz') of spheres and
that we may branch here to sphere Complexe and their 
geometry \protect{\cite{reye:1879}} as well. That's, 
however, beyond the current scope of presentation here 
(see e.g. \protect{\cite{dahm:GOL}} with respect to 
transfer principles) although there are 'tons of' very
interesting applications of this representation scheme
in physics. What we also don't want to discuss 
here in more detail is the interpretation of special 
relativity in terms of such sphere 'invariance' in 
different coordinate systems and with the additional 
constraint $x'=x$ and $y'=y$ or $\mathrm{d}x\mathrm{d}y=\mathrm{d}x'\mathrm{d}y'$
in the normal plane. Therefore, we need much deeper 
background with respect to sphere Complexe and Complex
geometry.} is the transition from typical 'light cone'
reps $x_{0}^{2}-\vec{x}^{\,2}=(ct)^{2}-\vec{x}^{\,2}=0$ 
to lines and the transformation of this constraint. 
Note already here that this framework can be applied 
also to point reps not on the light cone (or in 'momentum 
space' for 'massive particles' 'on the mass shell') 
by generalizing lines to 'Complexe', 'Gewinde' and 
null systems, 'Dynamen', 'Somen' or screws (see 
e.g.~\cite{study:1903} and references therein). Whereas
most usual treatments assume 'affine' point coordinates 
$x_\alpha$ in Minkowski's four-vector notation, we have
already pointed out (see \cite{dahm:MRST1} and \cite{dahm:QTS7})
that for same/equal 'time' $t$ in all four coordinates 
$x_\alpha$, the coordinate {\it value} $ct$ related to 
the coordinate $x_{0}$ has to be treated as infinity 
($\infty$) which can be done in (four) homogeneous/projective 
coordinates and the framework of projective geometry 
only\footnote{Please note, that this has to be discussed 
very carefully in terms of coordinate values and (binary) 
parameters, and care has to be taken in identifying 
homogeneous and inhomogeneous coordinates and their 
respective coordinate values/projection parameters.}.
The appropriate rep of space (point) coordinates, 
$x_{i}=v_{i}t$, in order to achieve an equally parametrized 
footing thus automatically introduces parameters
$\beta_{i}=v_{i}/c$ by using a (projective) Cayley-Klein 
metric when switching to inhomogeneous/affine (point) 
coordinates. The parameters $\beta_{i}$ which appear in 
physical transformations thus turn out as a reminiscence 
of line geometry while using inhomogeneous coordinates 
$x'_{i}\sim x_{i}/x_{0}$ in Euclidean descriptions of
3-dim real space. Although being -- in conjunction with 
points as basic space elements -- THE backdoor of Newtonian 
ideas and concepts within four-vector calculus, an 'overall'
(or absolute) parameter 'time' $t$ allows people to express 
dynamics by performing differentiation with respect to this
parameter while sticking to the point picture and its related
dynamical concepts, whereas part of the discussion can be 
mapped to velocities and their relations as is typically 
done in special relativity. But special relativity (see 
section \ref{sec:sgr}) can also be interpreted in terms 
of line and Complex geometry easily, and we use the 
individual/local times '$t$' and '$t'$' of two coordinate
systems only to select the respective subsets of lines 
out of all lines comprised within the geometrical setup.
So the task of (local) coordinates in a sense is to 
relate certain lines in a large 'line set' of a geometrical
setup. In other words, we can use 'times' to group and
sort lines or aggregations of lines (and related objects
like points, sections or higher order/class curves) within
the dynamical behaviour of the setup. Especially 'features'
like the invariance of normal planes (i.e. $x=x'$, 
$y=y'$ while translating along the $z$-axis) thus 
have straightforward geometrical background from 
Complex geometry and null systems.

Using a parameter $\epsilon^{2}=\pm 1,0$ to describe the 
respective non-Euclidean and Euclidean geometries, the transition 
of 'light cone' reps $\epsilon^{2}\vec{x}^{\,2}+x_{0}^{2}=0$ in 
terms of (4-dim) point coordinates $x$ (or $\epsilon^{2}u_{0}^{2}+\vec{\,u}^{2}=0$ 
in terms of (4-dim) plane coordinates $u$) into a line rep in 
terms of six related homogeneous line coordinates $\mathcal{X}$ 
is known to be performed by 
\begin{equation}
\label{eq:traegheit1}
\mathcal{X}_{01}^{2}+\mathcal{X}_{02}^{2}+\mathcal{X}_{03}^{2}+
\epsilon^{2}\left(\mathcal{X}_{12}^{2}+\mathcal{X}_{23}^{2}+\mathcal{X}_{31}^{2}\right)=0\,,
\end{equation}
or in the more symmetric form
\begin{equation}
\label{eq:traegheit2}
\frac{1}{\epsilon}\left(\mathcal{X}_{01}^{2}+\mathcal{X}_{02}^{2}+\mathcal{X}_{03}^{2}\right)+
\epsilon\left(\mathcal{X}_{12}^{2}+\mathcal{X}_{23}^{2}+\mathcal{X}_{31}^{2}\right)=0
\end{equation}
which simplifies Euclidean geometrical reps and discussions.
Whereas the general theory necessary for physics mounds at least
into the framework of quadratic line Complexe\footnote{German:
Quadratische Complexe}, here we want to mention only the fact 
that the lines of a Complex of $n$th degree, if they are incident
with one point (resp. they meet in one point) of $\mathbb{R}^{3}$, 
constitute a conic surface of $n$th order (\cite{plueckerNG:1868}, 
\S 2, p.~18), and the lines envelop a planar curve of $n$th 
class. 

{\it So quadratic Complexe constitute a ('light') cone of 
second order in 3-dim space}, meeting in one (or each) point 
as required by \cite{ehlers:1972} which we have physically 
associated with 'the photon' (see section~\ref{sec:context}
or \cite{dahm:MRST2}).

Thus, we can study associated planar conic sections of second
class which we can relate to (quadratic) invariants and energy,
however, the more striking feature of quadratic Complexe with
respect to relativistic requirements \cite{ehlers:1972} is 
their foundation in projective geometry which fulfil some 
requirements right from the beginning, and the overall 
integration between classical point/plane and line descriptions.
As soon as we interpret this 'light cone' (as usual) in terms 
of a 'metric' on point spaces and/or in {\it four metric} 
coordinates $x^{\mu}$, we have already introduced additional
physical identification or at least an additional dimension 
(i.e.~we would have to use {\it five homogeneous} coordinates,
see e.g.~\cite{klein:1928}, appendix~\S 5) in order to treat
absolute elements ('infinities'). From our viewpoint, it is 
much easier and much more consistent to understand the 
'light cone' as an (tangential part of an) absolute element
(or 'gauge surface') when switching from quadratic line/Complex
reps to (homogeneous) point reps $x$ already {\it included} 
in the projective description of 3-space. So based on a 
quadratic Complex (like the Pl{\"u}cker-Klein quadric), 
there is {\it no need} to impose additional geoemtrical
constraints and assumptions, nor is it necessary to impose 
or require an additional {\it axiomatic} framework of 'affine
geometry' like given and pursued e.g.~in \cite{weyl:1918}. 
Klein's 'Erlanger Programm' then provides a straightforward
guideline to fix invariant (geometrical) objects and find 
(restricted) transformation groups as linear subgroups of 
projective transformations. So using (quaternary) invariant
theory and approaching Euclidean (and differential) geometry
via 'affine geometry' and the Caley-Klein process, we can 
establish the known Minkowski metric without additional 
assumptions from quadratic Complexe and its related point
rep $x^{\mu}x_{\mu}=0$. Please note however the change 
in the {\it interpretation} of the coordinates $x^{\mu}$
which we've changed from the usual {\it metric/Euclidean}
interpretation to four {\it homogeneous} coordinates 
$x_{\alpha}$. Using quadratic Complexe, we control a 
superset {\it to derive} those features -- there is no
need to introduce them by hand, however, we have to 
perform Complex geometry. So first of all, the unifying
space element should be chosen as a linear Complex, and 
we have to relate our reasoning in 3-dim space to higher
order Complexe and calculation patterns in order to compare
to physics and extract principles.\\

The limit $\lim_{\epsilon\to 0}$ in eqns.~(\ref{eq:traegheit1})
or (\ref{eq:traegheit2}) towards Euclidean geometry has 
to be performed carefully. However, in this limit, we find 
from above the constraint 
$\mathcal{X}_{01}^{2}+\mathcal{X}_{02}^{2}+\mathcal{X}_{03}^{2}=0$
involving the $x_{0}$ coordinate(s) of the point rep(s). Besides
switching between \PL and Klein coordinates, we can complexify
further (individual) coordinates which changes the signature 
in line space (e.g.~in eq.~(\ref{eq:traegheit1})) as well as 
in point space. So in general, we have to discuss the related 
transformation groups SO($n$,$m$), $0\leq n,m\leq 6$ with 
$n+m=6$, or the related complex transformation groups 
SU($n$,$m$), $0\leq n,m\leq 4$ with $n+m=4$, or even 
quaternionic transformations like Sl(2,$\mathbb{H}$)
(or SU$*$(4), respectively). Dependent on the inertial 
index\footnote{German: Tr{\"a}gheitsindex} (or signature) 
of the quadratic form (\ref{eq:traegheit1}), we can of course
define linear reps and a 'metric' for a 'norm' being invariant 
under the respective SO($n$,$m$) symmetry group, $n+m=6$; 
SO(3,3) and SO(6) for \PL and Klein coordinates are well-known.
The general form
\[
a_{\alpha\beta}\mathcal{X}_{\alpha\beta}=0
\] 
defines a (linear) Complex $a_{\alpha\beta}$ in terms of line
coordinates $\mathcal{X}_{\alpha\beta}$, and dependent on the
\PL condition for the parameters $a_{\alpha\beta}$, we have 
to distinguish singular and regular Complexe, and apply the 
framework of Complex geometry and symplectic symmetries. 
Quadratic Complexe may be described\footnote{With respect to
rearrangements and discussion of uniqueness, see \cite{plueckerNG:1868}
or \cite{clebsch:1869}, eq.~(11) or the discussion following
eqns.~(15) and (16). With respect to the Pl{\"u}cker-Klein 
quadric $\Omega$ and the interpretation of (special) linear
Complexe in $P^{5}$, see \cite{klein:1872b} \S 1. We'll find
such invariants in section~\ref{sec:electrodynamics}.} by 
the general form $b_{\alpha\beta}\mathcal{X}^{2}_{\alpha\beta}=0$.

Last not least, in this context, we want to mention one more
aspect of our ongoing work in that \PL has associated Complexe
(resp. lines and axes) and especially Congruences of two or 
more Complexe to ellipsoids (see \cite{plueckerNG:1868}, 
'Erste Abtheilung', \S 3, p.~99ff, ibid.~\S 3, eqns.~(46)ff
or \cite{plueckerNG:1868}, 'Zweite Abtheilung', preface and 
main text) or various more general types of surfaces. There 
is indeed much older work \cite{pluecker:1838} where \PL 
defined such specialized ellipsoids in the context of Fresnel's 
wave theory, confocal surfaces and 'potential theory'. For 
us, this provides some geometrical background of the nowadays
usual mystification of the 'wave-particle dualism'. \PL (and 
other people at that time) knew well that, working with 
Complexe and (some of) their Congruences, one finds line 
reps (e.g.~axes) which have naturally associated ellipsoids
\cite{pluecker:1838}, and vice versa, and thus (strictly) 
spherical problems like Laplace or Schr{\"o}dinger
equations are special cases only. The separation denoted 
nowadays by this suggested 'dualism' is caused by describing
'point' particles by only half (i.e. the polar part) of 
the originally necessary line rep while playing games with
Euclidean/affine dynamics. So instead of mystifying the 
relation and interconnection of the two descriptions, one
should think in terms of lines and transfer principles.

Due to a line being a priori free in $\mathbb{R}^{3}$ (or $P^{3}$) 
to connect a point with an observer (i.e. always by its very
definition to connect at least two points), we can a priori 
handle (space-related) 'extension', different coordinate 
choices by investigating and/or transforming the fundamental 
tetrahedra and 'non-localities' especially of 'the 
photon'\footnote{The discussion of relating differential 
geometry to projective geometry has been a major topic for 
decades around the turn of the 19th to the 20th century.
However, the assumptions, specializations and drawbacks
introduced into differential geometry and calculuses seem
to be forgotten\ldots}. Tangential spaces are special cases 
of polar setups in conjunction with conics or surfaces which 
themselves can be treated by projective construction mechanisms 
and discussion of 'class' instead of 'order'. We can use the 
important apparatus of tangential \cite{plueckerNG:1868} and
tetrahedral Complexe (see e.g.~\cite{vonStaudt:1856}, \cite{reye:1866}),
moreover, we have a 'natural' definition of conjugation right
from geometry. Last not least, invariance of a line under 
transformations automatically provides (affine) translation
invariance when expressed in point coordinates, so with 
respect to the Poincar\'{e} group and contractions, we 
definitely have a well-defined geometrical framework which 
can be treated by lines or 'Gewinde' and geometrical limits
thereof \cite{zindler:1902}, \cite{study:1903}.

As an example, after having accepted line coordinates and 
line reps, one can easily apply incidence relations of
lines in (6-dim) line coordinates\footnote{German: 
Pl{\"u}ckersche Zeiger} $p_{A}$, $1\leq A\leq 6$,
and work e.g. with Klein coordinates\footnote{German: 
Kleinsche Zeiger} in order to relate equations like 
$\sum p_{\nu}p_{\nu+3}=0$ or $\sum x_{i}^{2}=0$ to the
framework of ruled surfaces (see \cite{zindler:1902}, 
Vol. 2, I \S 4). This facilitates a direct generalization
to Complex geometry.

\section{Physical Identifications}
\label{ch:physics}
As this is ongoing work, we'll mention briefly some aspects 
of identifying physics with such geometrical concepts.

\subsection{Electrodynamics}
\label{sec:electrodynamics}
We've argued already (see section~\ref{sec:context} or 
eq.~(\ref{eq:cosetrep})) within the framework of spontaneous
symmetry breaking (SSB) that we want to use a Goldstone 
identification of the (massless) photon in SU$*$(4)/USp(4)
in order to relate equivalence classes of velocities and 
the 'masslessness' of photons in common QFT frameworks.
The physical equivalence is the connection of velocity 
changes (in the coset) with photon emission 
('Bremsstrahlung'), and as a consequence, we relate
redefinitions of USp(4) Wigner-Weyl reps and especially 
the ground state to photon emission resp. (gauged) energy
changes. Although this is reasonable from the physical 
viewpoint in that we can relate (hard) observations to 
such models, the mathematical and physical formulations
using differential geometry at the one or other point 
look hazy. So people introduce 'velocity' 4-vectors 
$k^{\mu}$ 'on the light cone' and 'polarizations' 
$\epsilon^{\mu}$ with additional constraints which lead
to the one or other obscure explanation or philosophy.
In this context one can mention conditions like the 
'masslessness' (of 'particles') $k^{\mu}k_{\mu}=0$, 
the distinction of 'on-mass-shell' and 'off-mass-shell' 
behaviour (or 'virtual particles') in interaction processes
and 'gauge conditions' like $k^{\mu}\epsilon_{\mu}=0$ or 
even $\vec{k}\cdot\vec{\epsilon}=0$, i.e. 'orthogonality',
in conjunction with using normals of normals like with
$\vec{k}$, $\vec{E}$ and $\vec{B}$.

For us, the problem to determine a (vectorial) 'velocity' 
$k^{\mu}$ as a physically meaningful, linear dynamical 
object ends in front of the fact that light spreads out 
'on the light cone', i.e. by 'construction' on a second 
order (null) cone with the maximum (and for 'massive 
particles' unreachable) 'velocity' in order to transport
information. As such, we can honestly derive this spreading
from a construct with $\vec{k}$, $\vec{\epsilon}$ and the 
Poynting vector (i.e. from two 3-dim objects $\vec{E}$ and 
$\vec{B}$ respectively $\vec{H}$ related to the physical 
force $\vec{F}$) only -- while keeping in mind that in 
order to treat this type of infinity 'on the light cone', 
we have to use homogeneous coordinates! In this sense, we 
could use Klein's remark (see \cite{klein:diss}) that (for 
homogeneous coordinates!) the \PL condition 
$\sum p_{\nu}p_{\nu+3}=0$ is sufficient to define a line
(rep). The general way out of this problem is to use line
(or Complex) coordinates.

However, for us that's not really sufficient because we
are not only working with simple lines or linear Complexe, 
but also with quadratic ones (or at least with quadratic 
constraints using linear Complexe). Moreover, we know that 
electromagnetic {\it forces} related to $\vec{E}$ and $\vec{B}$ 
are to be described via the Lorentz force, and that in 
Hamiltonian (and also in Langrangian) formulations of 
dynamics we can start using $\vec{E}$ and $\vec{B}$ in 
terms of the antisymmetric field strength $F_{\mu\nu}$ 
-- although nowadays people prefer to use the description 
via the potential(s) $A^{\mu}$ and partial derivatives 
thereof, mostly as a trade-off to a Lorentz covariant 
description and differential reps. Whereas the rep of
$A$, as dependent of $k$ and $\epsilon$, can be naively
related to a line rep comprizing $\vec{k}$ and $\vec{\epsilon}$,
at the same time, we have to take care of the two normals
$\vec{E}$ and $\vec{B}$ and their dynamics, too.

Now a major point of discussion for us at the moment is a 
possible identification of the tensor $F_{\mu\nu}$ with a 
line rep (or a linear Complex). The 'tensor' character of
this object (with {\it two} indices) is caused formally 
only by Minkowski's four-vector formalism. We can ad hoc
associate the space components of $F_{ij}$ (the (Euclidean) 
vector components of $\vec{B}$ (or $\vec{H}$)) with the 
axial part of the 6-dim line rep and the components $F_{0i}$,
i.e. the components $\vec{E}_{i}$ (see e.g.~\cite{jackson:1983},
ch.~11), with its polar 3-dim part. Then the orthogonality
relation $\vec{E}\cdot\vec{B}=0$ may simply be interpreted
(see above) as the \PL constraint in eq.~(\ref{eq:plueckercondition})
to fulfil the line condition, although the association of 
a polar 3-dim vector rep with null-components in the face
of eq.~(\ref{eq:traegheit1}) and its Euclidean transition
seems to be not the best choice of identification. And yes,
we have to talk about six {\it homogeneous} line coordinates
which makes it difficult to interpret $\vec{E}$ and $\vec{B}$
directly in terms of physically observable or measurable
objects, but we have to keep in mind that also the charges
(as well as the masses) are only defined {\it in relation}
to another charge (or mass) as is known from Coulomb's
(and Newton's) law\footnote{This results also from Pl{\"u}cker's
identification of forces with respect to line reps, see 
references in \protect{\cite{plueckerNG:1868}}. So in
experiments we expect to see charge and/or mass relations
like reduced masses or physically observable combinations 
like $e/m$ only, which emphasizes the physical formulation 
by the Lorentz force when describing dynamics and (Lab) 
measurable 'accelerations'.}. The discussion of Lab 
measurement brings us back to discuss the introduction of 
(local) time '$t$' like in $\vec{F}=\tfrac{\mathrm{d}}{\mathrm{d}t}\vec{p}$ 
or $\vec{F}=m\vec{a}$ (indirectly).

Whereas we can use products like $F^{\mu\nu}F_{\mu\nu}$ to 
represent squares\footnote{And as such energies! According 
to our current understanding, that's the reason why the 
electromagnetic description works well on the classical as 
well as on the quantum level using Hamiltonian/Lagrangean 
formulation.}, our investigations especially in the context
of Complexe and (Complex) Congruences have started only. So 
as ongoing 'program', we have to map physical observations 
(i.e. objects and their dynamics!) to Complex geometry\footnote{This 
is in some parts not new but the problem is that science 
industry today uses (although limited in a lot of aspects)
all kinds of 'vectors' or linear reps and not line or even
projective geometry, and a lot of old knowledge is simply 
forgotten in favour of algebraic and analytic technicalities
around all kinds of linear vector spaces.}.

With respect to electrodynamics, the introduction of the 
'dual' 'tensor' $\mathcal{F}^{\alpha\beta}$ via 
$\tfrac{1}{2}\epsilon^{\alpha\beta\gamma\delta}F_{\gamma\delta}$
enhances the scenario and introduces further aspects into
the Complex representation\footnote{However, we do not want
to discuss transitions from ray to axis line coordinates 
and the related duality considerations of points and planes
in $\mathbb{R}^{3}$ here.}. From the viewpoint of line or 
Complex geometry, this 'new' object reflects advanced 
(algebraic) operations of a 6-dim line calculus in that 
we have to treat line incidences, i.e. 'products' of line
reps or parts thereof which resemble inner products or
'norms'. So the 'skew tensor' approach corresponds directly 
to (6-dim) line geometry, and products of (skew-symmetric) 
'tensors' are able to represent (6-dim) multiplications in
line coordinates, i.e. lines and incidence relations of 
lines. So at a first glance, line geometry works pretty 
well for electromagnetism in order to cover the four-vector
formalism. What is under construction (or 'open') at the 
time of writing, is the association between algebraical
and physical objects and a deeper understanding of line
Congruences\footnote{Especially also with respect to the 
identification of ray systems (German: 'Strah\-len\-sy\-ste\-me 
erster Ordnung und erster Classe').} as well as the 
physical meaning/identification of $\vec{E}$ and $\vec{B}$
versus $\vec{k}$ and $\vec{\epsilon}$.

There are indeed more sophisticated notions than the simple 
line concepts referenced so far. If we associate these 
3-dim 'field' reps to (linear) Complex parameters 
$a_{\alpha\beta}$ which (due to Cayley) can be interpreted
as line coordinates fulfilling the \PL condition, too, 
if $a_{\alpha\beta}p_{\alpha\beta}=0$ and $p_{\alpha\beta}$
are line coordinates of incident lines\footnote{German: 
Treffgeraden}, for (six) linear Complexe, a constraint 
formally similar to the \PL constraint can be formulated
as well to construct a quadratic Complex (see \cite{klein:diss},
Nr.~26).

A further extension of the Complex identification is based
on Complex geometry if we go back to the second order
surface given in eq.~(\ref{eq:traegheit1}) while choosing 
$\epsilon^{2}=-1$, and if we invoke polarity. Then the
two Complexe $C_{1}=\left(\mathcal{X}_{01}, \mathcal{X}_{02},
\mathcal{X}_{03}, \mathcal{X}_{23}, \mathcal{X}_{31}, 
\mathcal{X}_{12}\right)$ and $C_{2}=\left(-\mathcal{X}_{23},
-\mathcal{X}_{31}, -\mathcal{X}_{12}, \mathcal{X}_{01}, 
\mathcal{X}_{02}, \mathcal{X}_{03}\right)$ are polar with
respect to the surface. A simple calculation shows
\[
C_{1}\cdot C_{2}\equiv
-\mathcal{X}_{01}\mathcal{X}_{23}
-\mathcal{X}_{02}\mathcal{X}_{31}
-\mathcal{X}_{03}\mathcal{X}_{12}
+\mathcal{X}_{23}\mathcal{X}_{01}
+\mathcal{X}_{31}\mathcal{X}_{02}
+\mathcal{X}_{12}\mathcal{X}_{03}\equiv
0\,,
\]
so by eq.~(\ref{eq:lineincidence}) the Complexe $C_{1}$
and $C_{2}$ intersect, and we can start applying further
reasoning from Complex geometry and compare to physical 
observations.

\subsection{Special and General Relativity}
\label{sec:sgr}
\begin{figure}[h]
\includegraphics[scale=0.65]{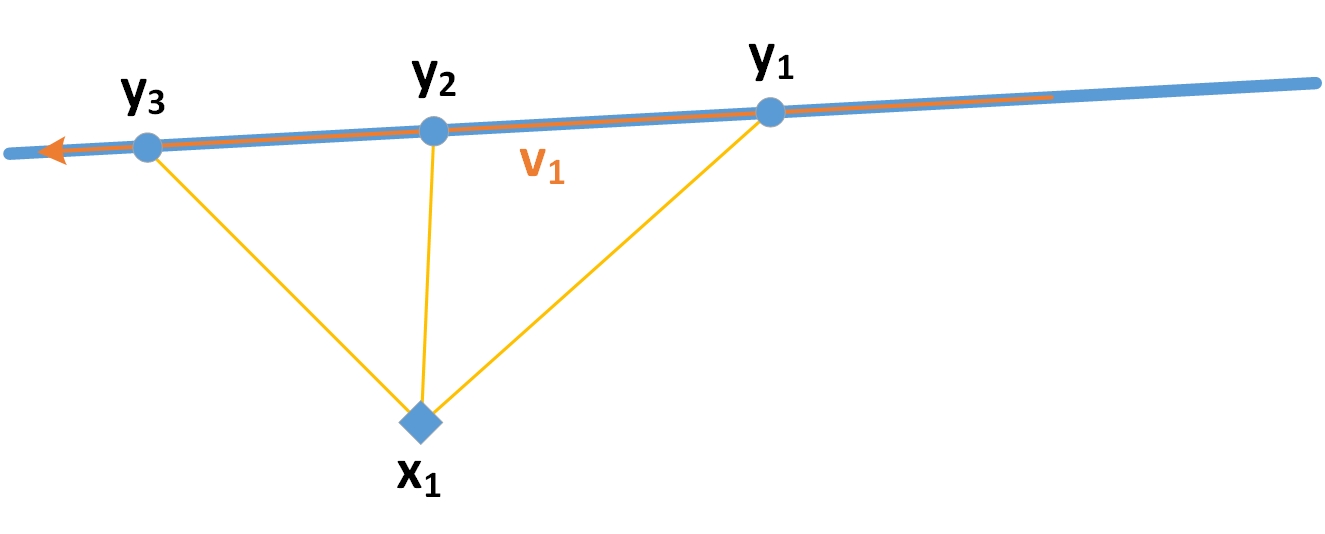}
\caption{Linear motion of points, observed by $x_{1}$.}
\label{fig:fig1}
\end{figure}

In order to extend what we have said above to 'relativistic'
physics, the simplest approach is to include observers 
right from beginning into the mathematical description.
This, too, is automatically provided using line geometry.
If we imagine for a moment the simplest scenario of an 
observer at rest watching a (non-accelerated) moving point 
(in some distance), then -- as time elapses -- we have at
a first glance the line of the (moving) point, i.e. a
collection of points at different time, of course, or
with different 'coordinates' parametrized by time. This
illustrates explicitly that if we use the 'physical'
information of the relative velocity of the two points
to parametrize the scenario the notion of time -- whether
from the observer's or the moving point's side -- is
needed to parametrize the individual coordinate notations
and definitions only. But moreover, we find a (planar) 
pencil\footnote{German: B{\"u}schel} of lines connecting 
the observer's point $x_{1}$ in space with points $y_{i}$
on the line re\-presenting the trajectory of the linearly
moving point (see Figure~\ref{fig:fig1}). 
This concept of an individually moving point, besides the
independent coordinate system of the line with points $y_{i}$
itself, allows to introduce a velocity {\it from the viewpoint
and in the coordinate system of the observer} which may be
synchronized easily assuming the Newtonian picture with overall
or absolute time throughout the complete description of the
system\footnote{Formally, the two subsystems have to be 
'synchronized' two a common coordinate system, i.e.~by exchange
and commitment of additional information. In Appendix \ref{app:HU}
we've discussed such a case for classical physics. Smilga 
\cite{smilga:2011} has presented a similar idea for the quantum
picture analyzing the tensor product of single-particle states.}.

If in addition, we allow for the observer to move freely\footnote{At 
first, we assume non-accelerated movements and skew/non-incident
lines of observer and point.}, the trajectory of the observer 
at $x_{1}$ is a line, too. Note, that we may immediately symmetrize
this picture by switching the r\^{o}les of point and observer, or
formally switching the coordinate system as known from classical 
physics and special relativity.
\begin{figure}[h]
\includegraphics[scale=0.65]{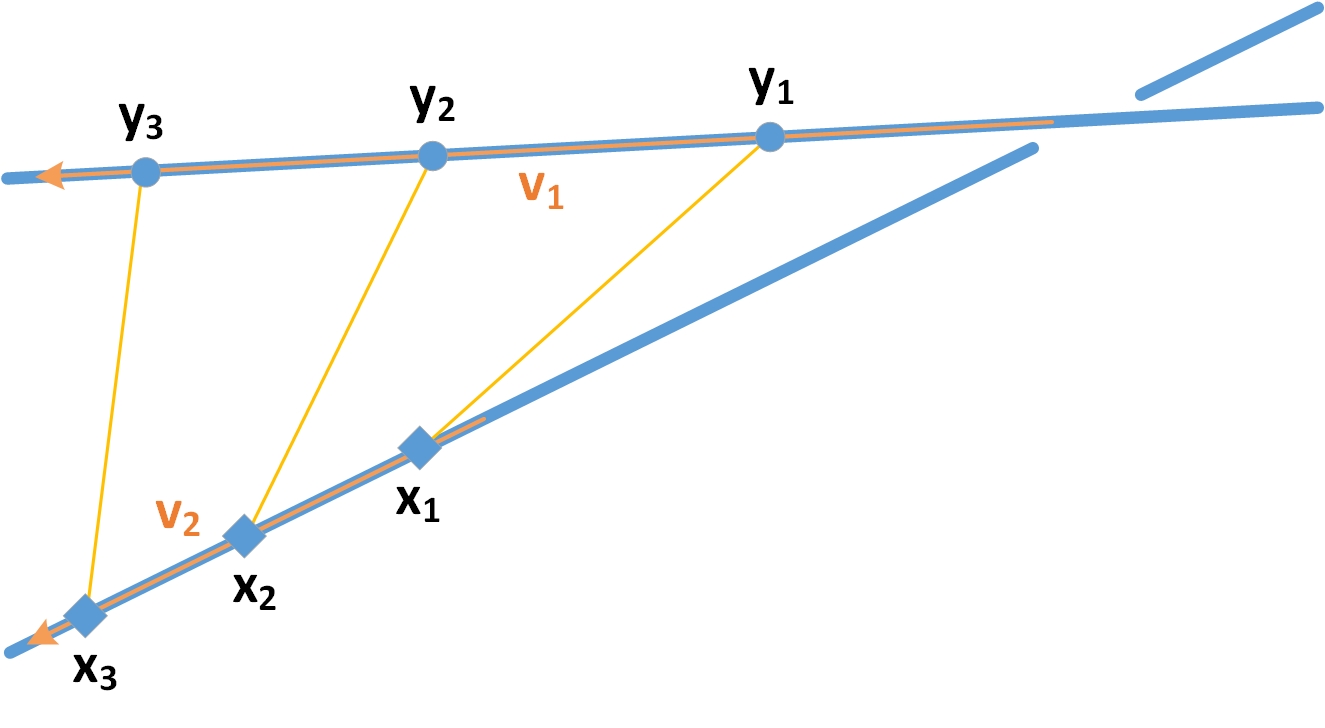}
\caption{Linear motion of points, observed by a linearly moving
observer $x_{i}$.}
\label{fig:fig2}
\end{figure}

If we now proceed as before by connecting points on the
two lines (see Figure \ref{fig:fig2}), it is the notion
and additional interpretation of 'moving points', where 
adjacently observed velocities $v_{i}$ connect 'times' 
to causality in the respective (local) coordinate systems
by ordering the points and their related description(s)
of physics. On the one hand, as such we can introduce 
and use individual (point) coordinate systems or apply
the typical reasoning of special relativity in terms of
point (or four-vector) coordinates. On the other hand, 
the picture of projective and especially line geometry 
offers two well-established frameworks to describe such
a setup much better by taking into account the individual
planar geometries of the two line pencils and the overall
picture of a ruled surface or a line Congruence (see 
Figure~\ref{fig:fig2}). In one approach, we may interpret
each of the lines as a special linear Complex, and identify
common lines within the superset of lines intersecting 
each of the two lines\footnote{German: Treffgeraden}. 
There is, however, a second possibility in that both lines
do {\it not} belong to the same linear Complex. Then, we 
may construct a regular linear Complex
$a_{\mu\nu}$ by $a_{\mu\nu}=\lambda p_{\mu\nu}+\rho q_{\mu\nu}$
where $p_{\mu\nu}$ and $q_{\mu\nu}$ are the line coordinates
of two skew lines $p$ and $q$.

Whereas we've given already some arguments and references
on line geometry and Complexe to justify the deep r\^{o}le
of Congruences, we also want to mention a background paper
with respect to the Figures \ref{fig:fig1} and \ref{fig:fig2}.
Whereas nowadays people tend to discuss 'time reversal',
often in the context of space symmetries and special relativity,
and mostly by means of Euclidean coordinate interpretations,
it seems helpful to recall Klein's paper \cite{klein:1872c} 
which reviews the (even at that time historical) discussions
and interpretations of complex numbers and measures within
the context of line geometry. Referring for details to Klein's
paper, he connects two complex points (a point and its conjugate
point) by a real line. So the two points can be seen as base 
points in order to define an involution on the real line.
Being quadratic, in order to resolve for the (complex) point,
one has to define an orientation of the line which has been
proposed by von Staudt. Klein resolves this apparently 
'artificial definition' by referring to projective definition
of measure (i.e.~the Cayley-Klein measure ('metric')) on the
line using the two points as base points. So the sign given 
by the definition of the measure distinguishes both sets of
base points, or the individual complex point, respectively. 
Using point pairs, he obtains von Staudt's interpretation,
and thus the orientation of the line. The final identification 
with the quadratic covariant $\Delta$ allows to represent
a complex point by three arbitrary points of a certain order.\\

A second major consideration regarding 'observers' is to
tag one plane in space as an 'absolute plane' and to shift
one point of the coordinate tetrahedron (or one of the two
observers above) to this absolute plane, i.e.~to $\infty$. 
We thus obtain affine geometry, and by an additional
assumption with respect to a circle in the absolute plane
(the 'infinite circle') Euclidean coordinates and parallelism.
If we associate this description with 'classical physics'
(where affine transformations leave the absolute plane 
invariant), then we may ask which physics is related to the
scenario in Figure~2 when both observers are close to each
other (and both are far from the absolute plane, or if no
'absolute plane' is present at all!). By analogy to a common
experimental setup, we suggest to identify this physical
picture to quantum theory where the 'observer' and 'point'
are both part of the interaction/the 'process'. Why? Thinking
of electron scattering on a target, $e(\cdot,\cdot)e'$, as 
related experimental setup, we have two major possibilities 
to let the first vertex $e\longrightarrow e\gamma$ happen. 
If the distance of this first vertex is far from the target, 
people usually assume outgoing real photons $\gamma$ and 
calculate the 'tagged photons' by standard relativistic 
kinematics and by measuring both electrons. So the second 
vertex of the reaction is going to happen 'far away' with 
(idealized) real photons at the target. If we shift, however,
the first vertex into the target by scattering the electrons
directly by or within the target, we have to work with 
'virtual photons', and the overall kinematics is governed
by the effective descriptions of $e(\mathrm{target},X)e'$-processes
with all known phenomenological implications. 

The difference of both pictures is that we move 'the observer'
(represented by the first vertex $e\longrightarrow e\gamma$
'to generate' the observation by a photon) close to the target
of the observation, and we want to have {\it only one} unified
framework to describe the process and the involved reps. 
That's why we want to use projective geometry to shift both
vertices freely, and we have to adopt our description of 
physics appropriately. In order to work with overall valid
reps, we thus emphasize line and Complex geometry in terms
of homogeneous coordinates of 3-dim space.

Besides being still free to use individual coordinate systems
related to each line (e.g.~in associating the six line 
coordinates to the sides/lines of a (fundamental) tetrahedron),
we can use in addition the two lines of the trajectories 
(moving observer and moving point) as opposite sides of a 
(third) tetrahedron and introduce 'overall' coordinates 
(with an additional unit point and (if necessary) absolute 
elements or by associating the framework of tetrahedral Complexe)
in order to establish a common description/coordinatization
of both systems. So we'll have to work out the algebraic 
relations of the respective six line coordinates of the two
individual line identifications used to describe the two 
individual coordinate identifications versus using a common
(fundamental) tetrahedron related to a parametrization by 
relative velocity and an abstract overall time which will 
result in identifying point sets of line incidences and 
harmonic ratios and relating them while respecting (some 
or all) properties of projective transformations. This 
reminds correlating one-particle rep descriptions in 
quantum field theory (QFT) in order to find common and
comparable physical behaviour like in Smilga's nice work
(see \cite{smilga:2011} or \cite{dahm:QTS7}). Moreover, 
we can 'collect' all the lines connecting the two 
trajectories (at different (individual) times and as 
such space points, of course) and describe them via line 
incidences\footnote{German: Treffgeraden} of both lines, 
or more general in a first step by singular 
Complexe\footnote{German: Treffgeradenkomplexe}, ray systems 
(see footnote before) and by appropriate Congruences. This 
can be done not only in Euclidean geometry but the framework 
of line geometry (because imbedded in projective geometry) 
is available also for all types of non-Euclidean geometries. 
The physical picture of such a description becomes transparent 
and clear if in mind we associate a 'light' source to both 
the moving point and the observer, and if we think in terms 
of rays being emitted by the point and by the observer, 
respectively. Nonlinear movements can then be described by 
e.g. higher order (or higher class) curves and surfaces, and 
projective geometry provides dimension formulas and a lot of 
further useful tools. Thus, the 'physics' or dynamics is 
directly related to the geometry of the respective curves 
and/or trajectories. As mentioned above, the breakdown to 
(squares of) line elements $\mathrm{d}s^{2}$ is possible 
in various ways and respecting/representing various geometries
and associated symmetry groups.

\section{Physics and Complexe -- a Programmatic Approach}
\label{ch:program}

Having gathered and summarized so far some geometrical aspects
as well as few Ans{\"{a}}tze in physics throughout the last 
section \ref{ch:physics}, the different 'types' of line sets
termed 'Complex', 'Congruence' and 'Configuration' by \PL 
\cite{plueckerNG:1868} need some context, especially in that
this geometry according to our opinion should be re-invoked
for analytical use. In section~\ref{ch:physics}, we have
summarized few aspects from electromagnetism and (special) 
relativity, and their relation to linear Complexe, which we
want to discuss in more detail in upcoming publications.
\PL himself has given a detailed account in \cite{pluecker:1865a}
on how to use Complexe, he has had suggested Dynamen, and
how to summarize the description of force systems of mechanics
\cite{pluecker:1865b}. This concept has had found a certain
closure and completion from the viewpoint of classical geometry
by Study's discussion of different geometries, Dynamen and
of Somen \cite{study:1903}.

Having seen the importance and usability of these concepts,
which seem to be directly and strongly connected to physical
terminology and geometrical descriptions, we feel the need
to approach this notion more programmatically, and as such
we want to propose and later on follow an identification
scheme based on line Complexe which we feel suited to unify
mechanics and various features of electrodynamics and special
relativity geometrically. Based on \cite{pluecker:1838} and
Pl{\"u}cker's considerations in optics, we have discussed 
in section~\ref{sec:plueckeroptics} an established geometric
possibility to relate lines or Complexe to sphere geometry, 
which we assume suitable to investigate quantum properties,
too. Thus, we feel prompted to propose the following 
Complex-based ordering scheme or 'program' in order to 
relate structures and content as far as we understand 
these relations now, starting by quadratic Complexe.

\subsection{Remarks on Geometry}
With respect to the general idea, and in order 'to resolve'
a natural 'chicken-and-egg' problem right at the beginning,
it is necessary to recall that -- starting from projective
geometry of 3-dim space -- points and planes are understood
as 3-dim 'objects', although represented by 4-dim homogeneous
coordinates $x_{\alpha}$ and $u_{\alpha}$ in $P^{3}$, and 
although people 'feel' a close relation to 3-dim space of
physical observations. If plane 'objects' are included,
a transfer principle ('duality') allows to symmetrize 
this picture as has been established already long ago 
by classical projective geometry (see e.g.~\cite{enriques:1903}
and references). The apparatus of this geometrical
approach relies on intersections and incidences being
preserved by projective transformations, which is close
to physical observation like e.g.~in the case of point-plane
incidence $x_{\alpha}u_{\alpha}=0$.

However, with respect to using line geometry, concepts 
are different. \PL has shown already in his early text
books (see references in \cite{plueckerNG:1868}) that 
line space is 4-dimensional, i.e.~each line can be 
represented by 4 parameters. Being derived only from 
Euclidean geometry, because equations in these four 
parameters do {\it not} preserve grade under projective
transformations, \PL in addition has introduced the 
determinant $\eta$ of his original four line coordinates
to establish covariance of the five inhomogeneous parameters,
thus preserving grade of the equations with respect to 
projective transformations. This concept may be embedded
in $P^{5}$ using six parameters (or line coordinates) 
as given in eq.~(\ref{eq:plueckervars}) with respect 
to their construction by point coordinates of $P^{3}$.
The \PL condition used above is necessary 'to reduce' 
4-dim line geometry to 3-dim geometry of real space, 
however, we may still use the established framework 
of projective geometry with respect to incidences,
intersections and transformations. In other words, 
although we use lines and 'generate' points by 
intersection of lines and span planes by joints of 
two incident lines, we are still consistent with
projective geometry, and we are moving on the solid
ground of Klein's 'Erlanger Programm'. 

Because this concept introduces a 'chicken-and-egg' problem
with respect to $P^{3}$ as the space of point-/plane-reps,
and $P^{5}$ as the space of Complex reps\footnote{We do not
want to guide or provoke a philosophical discussion of
the superiority of 3-dim, 4-dim or $n$-dim space here.
The same discussion has to take place if we use 5-dim
circles in the 2-dim plane or 9-dim spheres in 3-space,
etc. Due to the associated transfer principles, it is
obvious that we'll find {\it several} different, but 
equivalent reps for a physical object, and as well 
several descriptions of one and the same process in 
terms of the respective (suitable) reps.}, we have to 
find equivalence relations and transfer principles.
This may be established by eq.~(\ref{eq:plueckervars}) 
if we introduce the \PL condition, or the Pl{\"u}cker-Klein 
quadric in $P^{5}$, respectively. This establishes a 
quadratic constraint in $P^{5}$ which has to be fulfilled
for points in $P^{5}$ on the quadric in order to correspond
to lines in $P^{3}$. So in $P^{3}$, we have to take 
care of quadratic algebras and involutions additionally!
Moreover, the points\footnote{There are higher order 
elements in $P^{5}$ as well like 'lines', hyperplanes,
etc. which we do not discuss right now.} of $P^{5}$ 
are thus separated by the quadric, and accordingly, 
we have to distinguish regular and singular ('special')
Complexe. Because the relation to linear point coordinates
relies on the definition of the fundamental tetrahedron 
in $P^{3}$, we have to compare to lines in $P^{3}$, 
i.e.~ points in $P^{5}$ on the Pl{\"u}cker-Klein quadric.

So, the starting point here is not $P^{5}$, but while
representing the objects of $P^{5}$ in $P^{3}$, for us
it is natural to discuss quadratic and linear Complexe
in $P^{3}$ with respect to the very coordinate definition
of $P^{3}$ in terms of the fundamental tetrahedron.

\subsection{Tetrahedral Complex}
\label{sec:tetrahedralcomplex}
To work with line reps and Complexe in $P^{3}$, we thus
depart from the tetrahedral Complex and it's property, 
that the Complex lines intersect a (fundamental) tetrahedron
by constant harmonic ratio for all lines of the Complex
\footnote{A detailed exposition, based on von Staudt's 
work \cite{vonStaudt:1856}, can be found in \cite{reye:1866}.
However, there are fundamental relations to Binet's 
quadratic Complex of the normals of confocal $2^{nd}$ 
order surfaces \cite{pluecker:1865a}, to self-duality 
of the tetrahedral Complex and polar reciprocity
\cite{vonStaudt:1856}, to tangential systems of curves 
and surfaces \cite{plueckerNG:1868}, to projectively
related (3-dim) point spaces where the points are 
connected by lines which then form a tetrahedral 
Complex \cite{reye:1866}, to axes of conic sections
of a $2^{nd}$ order surface \cite{reye:1866}, to secants
of $3^{rd}$ order (spatial) curves \cite{reye:1866}, etc.}.
Besides self-duality of the Complex, we thus benefit
{\it at the same time} from invariance properties of
projective (point) transformations when working with
linear coordinates of points and planes, and of 
correlations (or duality, or general transfer principles).

So this picture allows to relate various well-known
descriptions and interpretations as soon as we have
developed the line/Complex part on a parallel footing
to classical affine and projective geometry in terms
of points, planes and their various generated, higher-order 
objects. So besides analytical expressions, we feel the
need to sketch briefly an accompanying 'program' in
section~\ref{sec:program} which we want to execute 
step by step in order not to get lost in a plethora
of analytical/geometrical details.

\subsection{Overview Complexe}
We've mentioned various examples of linear and quadratic
Complexe already above, referring to some physical 
associations and relations. 

As such, and with respect to known physical applications,
here we choose quadratic Complexe as general origin of 
our discussion, and try to drill down to linear Complexe
and to the standard geometry of 3-space, usually described
in (metric or affine) point coordinates. Note, that in the
spirit of the old geometers, we do not emphasize complex 
numbers and Riemannian geometry, but we see complex (and 
hypercomplex numbers) for a while as additional symbols 
to represent geometry appropriately on real spaces according
to the need to linearize the square '-1', e.g.~to treat 
non-intersecting situations or the closure of the projective
line which (formally) relates it to circles.

\subsection{Quadratic Complexe}
\label{sec:program}
We've commented on the central r\^{o}le of the tetrahedral
Complex and its relation to the fundamental coordinate 
tetrahedron above. We may add that for special tetrahedra
(having e.g.~their four vertices on a sphere), we can discuss
polar relationships and include the center of the sphere as
unit point. This simplifies the coordinate description of
spheres in that we may omit weights and end with a sum of
squared coordinates.
The tetrahedron is related to tetrahedral Complexes because
all lines in (3-dim) space constitute fixed harmonic ratios
with respect to the points of intersecting the four tetrahedral
planes. So we may use these ratios to group lines in space.

This suggest a certain program to follow:
\begin{enumerate}
\item As mentioned in section~\ref{sec:tetrahedralcomplex}, 
considering 'W{\"u}rfe' \cite{vonStaudt:1856} and their 
invariance properties with respect to projective
transformations, we have found a base point or departure
of invariance discussions because projective transformations
(i.e.~the basic tool of Klein's 'Erlanger Programm') can be
directly related to invariance groups. Here, we can attach
the discussion of discrete and continuous/Lie groups, and
we can also attach the construction scheme of projective 
geometry, departing from 1$^{st}$ order/class objects, and
constructing higher order/class objects while preserving 
invariance of incidences. Typical questions comprise 
coordinate systems and calculus with respect to the different
scenarios/objects.

\item If we depart from projective geometry and, in a next 
step, fix additional geometrical structures as 'invariants'
(or 'absolute') objects, on the one hand, we follow precisely 
Klein's 'Erlanger Programm'. On the other hand, we have already
discussed above the case of an absolute plane which leads
to 'affine' coordinates, and Klein's introduction of the 
metric in the Euclidean case as an 'absolute circle' in the
'absolute plane'. Due to several reasons, we want to extend
this approach as we have argued already in section 
\ref{sec:plueckeroptics}. There, we have associated the 
typical 'metric' in Minkowski space with an invariance 
of a quadratic Complex\footnote{\ldots and the special 
choice of coordinates in point space from above.}, and 
we'll have to investigate more consequences of parameter
choices of the quadratic Complex as well as further 
possibilities of different absolute elements by means 
of quadratic or linear Complexe. In section~\ref{sec:metricNEW},
we'll discuss also a related change in the very coordinate
definition which yields interesting results with respect
to relativity and typical quantum formulations.

\item Having fixed a $2^{nd}$ order surface in $P^{3}$,
we may benefit from the Cayley-Klein approach to metrics,
i.e.~the metric is {\it not} given by god (or a scientist),
but it is strictly related to a geometrical setup and 
some of the assumptions (see eq.~(\ref{eq:traegheit1})
or (\ref{eq:traegheit2})) above. Then, of course, we may
fall back to purely analytical discussions (having to 
take care about the coordinates, however), and proceed
with point and appropriate differential geometry. From
the viewpoint of line geometry which we want to pursue,
it is however helpful to investigate the (quadratic) 
tangential Complex \cite{plueckerNG:1868} and its 
properties. Besides the relations between such types 
of Complexe and $2^{nd}$ order surfaces, we want to 
recall also the normals of confocal surfaces which 
constitute a $2^{nd}$ order cone \cite{plueckerNG:1868},
and for our own considerations later on, we want to 
recall the generation of $2^{nd}$ order surfaces by 
lines. So with respect to tangential considerations 
(where we nowadays discuss dynamics and Lie theory),
it is obvious that we may use as well an approach by
generating lines and polar systems in order to cover
the tangential discussion not only in this singular 
point of the tangent space, but line geometry and 
Complexe allow global calculations and definite, and
controllable, reductions schemes to the tangential
case.

\item So one of the last steps which are needed to
complete such a program is the reduction of quadratic
to linear Complexe, or in other words, to find consistent 
analytic reps in point/plane as well as in line space
to represent square roots. So whereas this concept is 
already well-known from Dirac's approach to quantum 
theory to resolve the Minkowski 'norm' into two conjugate
linear reps, however, to our opinion the 6-dim line 
(and momentum) reps are much better suited, because 
forces and force systems have to be described by 
Complexe (see \cite{pluecker:1865b}, \cite{plueckerNG:1868}
and references). As such both sources, \PL \cite{plueckerNG:1868}, 
and Klein's reprise \cite{klein:1872b}, are relevant
from the physical viewpoint in that Klein (\cite{klein:1872b}, \S 2)
with respect to an arbitrary Complex line and the 
quadric $\Omega$ in $P^{5}$ discusses the ambiguity
of the related linear tangential Complex, and he 
points out the relevance of a special linear Complex
as well as a special Congruence. Mathematically, we
may use Clebsch's discussion \cite{clebsch:1869} to 
proceed to Complex calculus, and quaternary forms and
invariants. We want to apply this discussion to matter
fields, and to photons coupling to matter, as well as the 
typical Lagrangean descriptions of QED and gauge fields.
Due to the identification of the photon with a special
linear Complex in section~\ref{sec:electrodynamics}, 
we have to investigate the role of linear special
Complexe as well as the role of regular Complexe in
both of the geometrical contexts of $P^{3}$ and $P^{5}$.

\item So last not least, we have to focus this 'program'
to usual Hamiltonian or Lagrangean formulations in order
to finally discuss equations of motion and conserved 
quantities. In other words, the connection can be found
by the usual invariant theory of Hamiltonian/Lagrangean
formalisms and its relation to quaternary invariant 
theory of Complexe in terms of coordinates in $P^{3}$,
or even directly of forms and invariant theory in $P^{5}$.
The variation with respect to (general or restricted)
projective transformations can be formulated in terms
of $\delta\mathcal{L}=0$, if the Lagrangean can be
expressed in irreps of either point/plane combinations
and forms in $P^{3}$, or 6-dim reps and invariants of
lines or linear Complexe. So we'll have to investigate
invariants of QED type $F_{\mu\nu}F^{\mu\nu}$ or
Yang-Mills type $F^{a}_{\mu\nu}F^{a\,\mu\nu}$, $1\leq 
a\leq 3$, in the Lagrangean framework, and we can 
compare to Klein's discussion of six linear fundamental
Complexe which span $P^{5}$. Besides the identification
of the individual geometrical objects, the major problem
will be to find the correct coordinate reps with respect
to the symmetry groups SO($n$,$m$), $0\leq n,m\leq 6$, 
where $n+m=6$, and the groups SU($n$,$m$), $0\leq n,m\leq 4$,
where $n+m=4$, and their respective physical interpretations.
\end{enumerate}

\subsection{'The Metric' Revisited}
\label{sec:metricNEW}
In order to gain more control on coordinates and especially 
the metric as discussed in section~\ref{sec:themetric}, it is
noteworthy to recall some old geometry. Usual geometrical and
physical considerations often assume -- at least intrinsically
-- rectangular coordinates and differentiability like $\mathrm{d}x_{i}$
(or their two relevant ratios $p$ and $q$, respectively) when 
working with 3-dim space and describing physics. In order to 
gain more control, we do not start from Cartesian descriptions
or Weyl's axiomatization of 'affine geometry', but it seems 
helpful to recall that the 'extension' of Cartesian/Euclidean
coordinates $x'_{i}$ to affine coordinates contains the assumptions
cited above, i.e.~the {\it additional} assumption of an invariant
{\it plane} 'at infinity', and a polar system, the 'absolute 
circle' in this 'absolute plane', to handle parallelism and 
orthogonality. Please remember during all our discussion, that
the concept of 'absolute' or 'ideal' elements historically has
been introduced to unify the analytical description of geometry,
e.g.~of (planar) lines intersecting {\it always} in {\it one} 
planar point!

Analytically, in usual terms one introduces four 'homogeneous 
coordinates' $x_{\alpha}$ of 3-dim space, $0\leq\alpha\leq 3$,
by
\begin{equation}
\label{eq:threeaffinecoordinates}
x'_{i}=\frac{x_{i}}{x_{0}}\,, 1\leq i\leq 3\,,
\end{equation}
or from a more general and complete view
\begin{equation}
\label{eq:fouraffinecoordinates}
x'_{\alpha}=\frac{x_{\alpha}}{x_{0}}\,\sim\,
\left(x'_{i},1\right)\,, 0\leq\alpha\leq 3\,,
\end{equation}
thus giving rise to the usual 'reduction scheme' to the primed 
Euclidean 'point' coordinates $x'_{i}$, used e.g.~in 
eq.~(\ref{eq:linecoordpluecker}). Here, $x_{0}=0$ describes 
the coordinatization of the 'absolute plane' by homogeneous 
coordinates, and the coordinates $x'_{i}$ as quotients of 
homogeneous coordinates are associated with the three Euclidean 
coordinates. So intrinsically (although people usually suppress
the fourth coordinate $\sim 1$), the Euclidean coordinates 
(\ref{eq:fouraffinecoordinates}) remember their relation to
the absolute plane of the affine picture. Details can be found
e.g.~in \cite{enriques:1903}. So the geometrical and synthetical
identification of the 'absolute plane' in projective geometry
finds its analytical counterpart in an intrinsic divergence
of the Euclidean coordinates for $\lim_{x_{0}\to 0}$. Taking
this limit $\lim_{x_{0}\to 0}$ on a second order sphere 
$S=\sum_{\alpha} x_{\alpha}^{2}=0$, $S$ yields $\sum_{i} x_{i}^{2}=0$,
i.e.~we 'find' a circle in the 'absolute plane' $x_{0}=0$ 
constituted by the three remaining homogeneous coordinates
$x_{i}$ which thus can serve as general planar coordinates.

Now, this coordinate definition, and the 'affine' notion, both
appear to be consistent as long as we regard transformations 
which leave this plane invariant, i.e.~transformations of points
and planes which do not alter the $x_{0}$-coordinate, as is 
usually assumed to be a property of 'affine' transformations. 

However, care has to be taken already with Lorentz transformations.
While in usual 'Euclidean' notation, the plane normal\footnote{Classical
orthogonality involves the polar system at $\infty$, i.e.~in the 
absolute plane.} to the velocity is left invariant by Lorentz 
transformations, the coordinates of the velocity direction and
'time' mix. On the other hand, everybody 'knows' that (special)
Lorentz transformations leave 'the norm' $x_{0}^{2}-\vec{x}^{\,2}$
invariant\footnote{For several reasons, we'll use the homogeneous
interpretation of these coordinates, and 'the norm' 
$x_{0}^{2}+\epsilon^{2}\vec{x}^{\,2}$ to comply with 
eqns.~(\ref{eq:traegheit1}) or (\ref{eq:traegheit2}),
i.e.~$g^{\mu\nu}\cong\mathrm{diag}(1,-1,-1,-1)$.} which may be 
interpreted as a cone (according to rank and signature of the 
spatial form), or in general as a $2^{nd}$ order surface.

Now, in order 'to produce' geometrically the same behaviour 
of infinite coordinates $x'_{i}$ like in the affine case 
(\ref{eq:fouraffinecoordinates}) for $x_{0}\to 0$, we necessarily
have to switch to a more general and 'more symmetrical' 
coordinate definition $y'_{\alpha}$ in 3-dim space,
\begin{equation}
\label{eq:neuekoordinaten}
y'_{\alpha}=\frac{x_{\alpha}}{\sqrt{f(x_{\alpha})}}\,,
0\leq\alpha\leq 3\,,\quad
f(x_{\alpha})=x_{0}^{2}-\vec{x}^{\,2}\,.
\end{equation}
So with respect to this 'new' set of coordinates $y'_{\alpha}$,
we can check their behaviour with respect to homogeneity 
(i.e.~$x_{\alpha}\to\lambda x_{\alpha}, \lambda\in\mathbb{R}$).
We obtain $f(\lambda x_{\alpha})=\lambda^{2}f(x_{\alpha})$, or
$\sqrt{f(\lambda x_{\alpha})}=\pm\lambda\sqrt{f(x_{\alpha})}$,
so the linear coordinate definition $y'_{\alpha}$ in (\ref{eq:neuekoordinaten})
as a quotient is independent of $\lambda$, and thus they show
the same behaviour as the 'original' Euclidean coordinates 
$x'_{\alpha}$ in eq.~(\ref{eq:fouraffinecoordinates}). If we
now build the expression ${y'_{0}}^{2}-{y'_{i}}^{2}$ while
using eq.~(\ref{eq:neuekoordinaten}), this expression in the 
four new coordinates $y'_{\alpha}$ yields
\begin{equation}
\label{eq:absolutequadric}
{y'_{0}}^{2}-{y'_{i}}^{2}=\frac{x_{\alpha}x^{\alpha}}{f(x_{\alpha})}
\equiv\frac{f(x_{\alpha})}{f(x_{\alpha})}=1\,,
\end{equation}
and it is even {\it independent} of the (original) homogeneous
coordinates $x_{\alpha}$ of 3-dim space at all as long as 
$x_{\alpha}$ in eq.~(\ref{eq:neuekoordinaten}) is acted upon 
with SO(3,1) transformations, i.e.~the expression 
$x_{\alpha}g^{\alpha\beta}x_{\beta}$ is preserved. We'll have
to discuss parallels to the absolute circle $\sum_{i} x_{i}^{2}=0$
from above elsewhere. Using the first equality in (\ref{eq:absolutequadric}),
it is obvious that this relation is independent with respect 
to $x_{\alpha}\to\lambda x_{\alpha}$, so the homogeneity of
the four coordinates $x_{\alpha}$ doesn't influence the quadric
in the new coordinates $y'_{\alpha}$, however, the {\it linear}
coordinates individually (and symmetrically) tend to $\infty$
as the original quadric approaches the 'light cone'. The
most important difference between $x'_{\alpha}$ in 
eq.~(\ref{eq:fouraffinecoordinates}) and $y'_{\alpha}$ in
eq.~(\ref{eq:neuekoordinaten}), however, is the suitability
to define valid coordinates as long as we transform the 
{\it linear} reps $x_{\alpha}$ by {\it all} SO(1,3) 
transformations. Note, that this is in general {\it not 
the case} for 'affine coordinates' according to 
eq.~(\ref{eq:fouraffinecoordinates}) due to $x_{0}$ appearing
linearly in the quotient.

With this 'new' coordinate definition $y'_{i}$, we take care of
the symmetry that the full group of special relativity defines 
appropriate limits with respect to one and the same 'absolute 
element' $f(x_{\alpha})=0$, the invariant sphere, and not an 
non-invariant plane $x_{0}=0$. This has, of course, enormous 
consequences which we'll have to discuss elsewhere. For now, 
it is sufficient to discuss the special case of affine geometry
by the equation
\[
\sqrt{f(x_{\alpha})}=\sqrt{x_{0}^{2}-\vec{x}^{\,2}}
=\sqrt{x_{0}^{2}}\sqrt{1-\frac{\vec{x}^{\,2}}{x_{0}^{2}}}
=x_{0}\sqrt{1-\frac{\vec{x}^{\,2}}{x_{0}^{2}}}
\]
So the 'new' set of coordinates $y'_{i}$ in eq.~(\ref{eq:neuekoordinaten})
comprises the standard Euclidean coordinates up to a well-known
factor,
\begin{equation}
\label{eq:lorentzfactor}
y'_{i}
=\frac{x_{i}}{x_{0}\sqrt{1-\frac{\vec{x}^{\,2}}{x_{0}^{2}}}}
=\frac{x_{i}}{x_{0}}\frac{1}{\sqrt{1-\frac{\vec{x}^{\,2}}{x_{0}^{2}}}}
=x'_{i}\frac{1}{\sqrt{1-\frac{\vec{x}^{\,2}}{x_{0}^{2}}}}\,,
\end{equation}
where we recover the old Euclidean, 'affinely constructed'
coordinates $x'_{i}$ of eq.~(\ref{eq:threeaffinecoordinates}).
If we consider only special relativity, we may assume the 
typical identification of a light cone (or light 'sphere') 
by $x_{0}=ct$, and while considering uniform linear motion
in the coordinate projections $x_{i}$ by introducing appropriate
velocity projections $x_{i}=v_{i}t$ per same 'overall' time 
interval $t$. So the fraction most right in eq.~(\ref{eq:lorentzfactor})
above is based on the ratio of velocities, and equals to
$\gamma=\frac{1}{\sqrt{1-\beta^{2}}}$, the Lorentz factor.
So eq.~(\ref{eq:lorentzfactor}) reads as $y'_{i}=\gamma\,x'_{i}$.
The denominator can be decomposed further by 
$\sqrt{1-\beta^{2}}=\sqrt{1+\beta}\sqrt{1-\beta}$, or if we
switch back to homogeneous coordinates, according to
$\sqrt{x_{0}^{2}-\vec{x}^{\,2}}
=\sqrt{x_{0}+\|\vec{x}\|}\sqrt{x_{0}-\|\vec{x}\|}$ which 
both suggest to investigate higher order/rational curves
and elliptic functions (or even integer squares, or triangular
numbers) later on.

\section{Outlook}
\label{ch:outlook}
Having in mind how we have associated (physical) light with
'light cones' above, we have additional possibilities on a 
linear representation level to generalize lines (i.e. 
singular Complexe) to general linear (and higher degree)
Complexe and, moreover, we can investigate their relation
to 'massive' reps 'on' and 'off' the mass shell. So what 
is open today is an a priori explanation of the Hamiltonian 
structure (and as such the energy) of being quadratic 
in line coordinates\footnote{Right now, we can conjecture 
only that this quadratic form is related to the fact that
the line representation of $\mathbb{R}^{3}$ is four-dimensional
and we need a (quadratic) constraint to eliminate one 
degree of freedom/dimension. There are further (quadratic) 
explanation possibilities originating from tangential 
and tetrahedral Complexe or from the $M_{4}^{2}$ above.
The possibility that a second order description of energy
itself is an approximation only and that we have to treat
this question based on general homogeneous functions is 
beyond scope at this time of writing.}. So the generalization
(and also our ongoing program if we think on how to approach 
general relativity) is twofold: We can extend the use and 
application of Complexe and Complex geometry, and we can 
investigate their various constraints with respective 
mappings to physics.

Because differential geometry (by using forms linearly) 
a priori reflects only (polar) parts of line reps and affine
behaviour, we are convinced to find additional energy-momentum
contributions to $T^{\mu\nu}$ (in four-vector notation)
by simply taking elements of line or Complex reps (e.g.
moments) into account or even more sophisticated mechanisms
of line or Complex geometry. Moreover, we see differential 
forms, Pfaffian equations (one-forms) and Lie theory as 
subsidiary concepts of line geometry only in their respective 
(inhomogeneous) 'time' dependent limits. Using line geometry,
the inclusion of observers is a priori guaranteed by the 
formalism, i.e. we do not have 'observer-free' physics 
as nowadays usual, and there is no need to speculate on 
'non-localities'. Last not least, with respect to dimensions
of transformation groups and concerning reality conditions 
so far, we want to mention the possibility that in choosing
a 'correct' set of coordinate reps we may associate the 
15-dim resp. 16-dim transformation groups with projective 
transformations mapping (linear) Complexe to Complexe and 
the 10-dim subgroup mapping null lines\footnote{German: 
Nullgerade} to null lines. However, that's an open issue 
right now and has to be proven formally which we want 
to pursue by executing the 'program' sketched in 
section~\ref{sec:program}. 

At the time of writing, we see various still 'competing' 
possibilities\footnote{Here, we want to pinpoint once more 
\cite{enriques:1903}, appendix II, this time in the context 
of mapping his two abstract spaces $S$ and $S'$ by collinear
mappings.} to work with 5-dim $p$ (or $\exp p$) and identify
the space physically (better: dynamically!), and we feel the
need also to discuss the double 15-dim (automorphic) collineations
of $M_{4}^{2}$ with respect to physics and real/complex 
descriptions much deeper in those contexts. This is especially
interesting when starting from Hamiltonian formalism while
using/identifying line coordinates and assuming the quadratic
structure as originating from $M_{4}^{2}$ as the fundamental
form (see \cite{klein:1872a}). We have given above already one
application of polarity, however, we have to investigate the
physical consequences in much more detail. We have mentioned
as well tangential Complexe and Congruences which we have
to arrange versus the current concept of affine connections
(see e.g.~\cite{klein:1872b}).

Another open problem is the identification of Complex-related 
numbers and/or constants versus reps and rep dimensions in 
the Lie group/algebra related approach, i.e. there are 15 
constants mapping the \PL constraint linearly onto itself, 
there exist polar systems depending on 20 constants (and series
thereof as well) which map (arbitrary) lines to linear Complexe
and vice versa, etc.

We are convinced \cite{dahm:QTS7} that various low-dimensional
Lie groups and algebras, especially su(2)$\oplus$u(1), occuring
in various applications of QFT are artefacts of certain aspects
of line (and projective) geometry of $\mathbb{R}^{3}$ which
emerge by taking and generalizing certain analytic and
algebraic aspects of line (and Complex) reps only in terms
of individual 'calculuses' and 'rules'. As such we see 
Pl{\"u}cker's $M_{4}^{2}$ and the twofold 15-dim automorphisms
in a central r\^{o}le, governed however by the rules of
projective geometry. In this context, there are lots of 
further deep geometrical connections to other topics like
Kummer's surface, Darboux's 5-dim reps of confocal 
cyclids\footnote{German: Konfokale Zykliden}, Pasch's 
sphere Complexe and their geometry or to rep dimensions
occuring both in line/Complex geometry and physical/QFT
rep identifications which we have to work on.

\subsection{Remarks}
Having been asked to contribute to a topical collection of 
AACA, in memoriam of Waldyr Rodrigues, for me it is a real
honour to contribute, and to dedicate this work to such a 
great mathematical physicist. Having been a great scientist
who left a lasting personal impression, and a real maker 
with respect to his university environment and to the AACA
community, he is and will be very much missed!

With respect to the circumstances and the focus on relativity
and field theory, I've decided to contribute the original, 
extended and so far unpublished version of the ICCA 10-contribution
in Tartu, Estonia, 2014, where our last personal meeting took 
place. The content presents geometrical aspects of field theory,
electromagnetism and relativity from a projective point of
view, and tries to fit to Waldyr's work and interest in those
topics.

Since then, we've published several parts of the series related 
to aspects and steps discussed in section~\ref{ch:program}, and 
especially subsection~\ref{sec:program}. So while executing 
this program, we have included Cartan's spinor calculus, based
on Study's work, by relating it to Lie's transfer of line 
and Complex geometry to spheres \cite{dahm:2017a}. Moreover, 
we have shown \cite{dahm:2017b} so far that Minkowski's 4-vector
calculus basically relates to (linear) Complex and projective
geometry. In both cases, work on further identifications and 
calculations is ongoing.

\appendix

\section{A Projective Setup}
\label{app:HU}
\begin{figure}[h]
\includegraphics[scale=0.55]{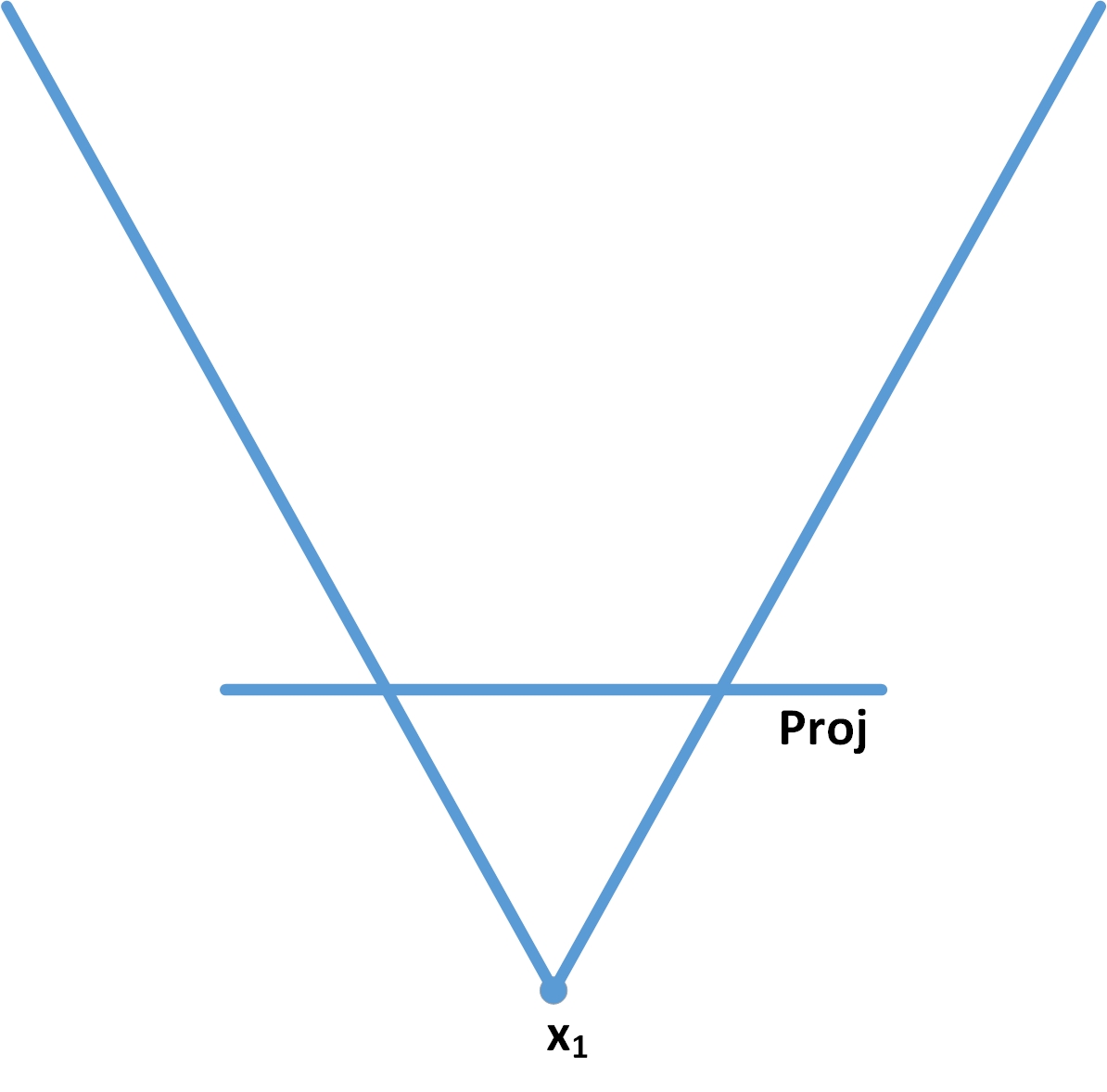}
\caption{Geometrical setup of observer and observation.}
\label{fig:app0}
\end{figure}

As an interesting and self-evident scenario we want to discuss
the scenario depicted in Figure~\ref{fig:app0} with respect 
to physical aspects.
In this setup, we assume that the observer located at $x_{1}$
can observe a (planar\footnote{The same arguments will hold
when enhancing the setup of the scenario to 3-dim space.}) 
physical process only by means of a projection to the line
'Proj', and the measurement of the projection on 'Proj' takes
place at equal time intervals $\Delta t$. So the individual 
time values of the observations are associated to positions
$x_{i}$ on 'Proj', and we can also introduce the concept of
velocities to quantify changes (or dynamics) with respect to
different observations and the interval $\Delta t$ from the
viewpoint of the observer. 

The observer 'lives' on one side of 'Proj', the physical 
process takes place on the other side. For the sake of 
simplicity of the description, we assume moreover, that 
the observer can watch the process only within the view 
limited by the two lines in Figure~\ref{fig:app0}.

\begin{figure}[h]
\includegraphics[scale=0.55]{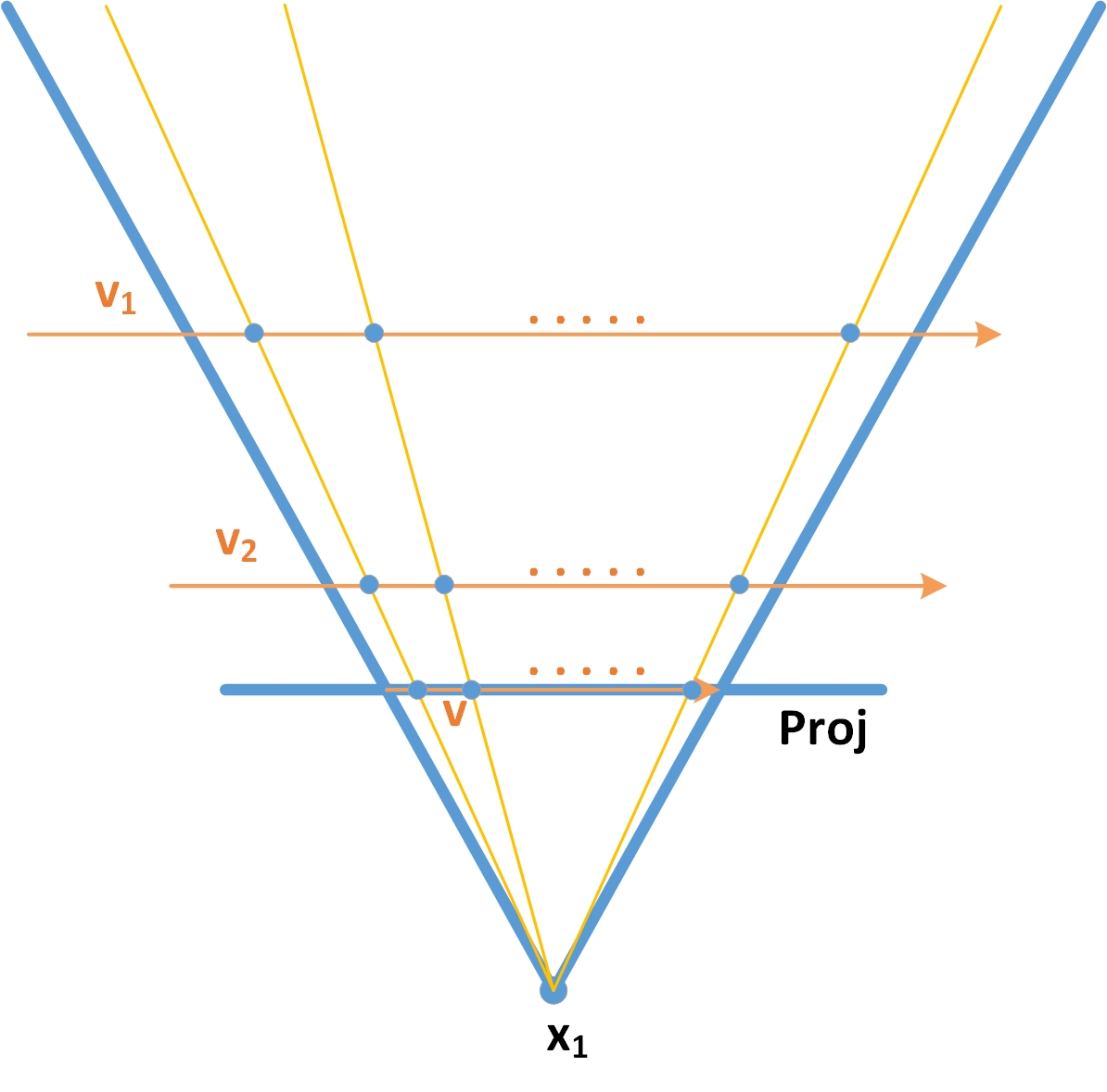}
\caption{Observer and observed parallel/linear velocities.}
\label{fig:app1}
\end{figure}

Now, if a point moves with {\it slow\footnote{We do not 
want to include relativity here.}constant} velocity 
$\vec{v}$ infinitesimally close and parallel to the 
projection line 'Proj' (see Figure~\ref{fig:app1}),
we can determine its velocity by, e.g., measuring its
location (more precisely, the location of the projection
on the line 'Proj') at certain times and calculating 
the velocity classically. If we draw virtual projection
lines from the observer to the observed points on the 
projection line (where we measure space versus time), 
we obtain first of all a pencil of lines with center 
$x_{1}$. The description of the velocity will correspond
in this case to the real physical velocity of the point.

However, if there are further movements of points {\it 
parallel} to the original process described by $\vec{v}$
and with constant velocities $v_{1}$ and $v_{2}$ like in
Figure~\ref{fig:app1}, they cannot be distinguished
from the first process as long as the velocities are
higher and meet at the same time of measurement the 
projection lines (of our observation/our observer). 
It is obvious that in order to establish such an 
identification, the velocities have {\it to increase}
dependent on the distance from the projection line/the
observer (which, however, is beyond the scope the observer
at $x_{1}$ can detect or know, having only data from 
measurements on the projection line 'Proj'. So the
absolute value $\|\vec{v}\|$ is a lower bound on the
velocity, and the observer -- without additional 
information -- is not able to determine the exact 
value of the 'real'/physical velocity of the point.

In other words, already in this simple setup, we find
equivalence classes of velocities as a function of 
distance (or space), and in addition, we find a lower
bound on $\|\vec{v}\|$ dependent on the position (or
distance) of the projection line from the observer.
These uncertainties are already apparent in classical
physics, and the only possibility to resolve this scenario
and perform complete and correct analytical calculations
is to gain {\it additional} knowledge on the setup, 
here on the additional distance information in order
to determine the exact $y$-coordinate of the moving 
point if we assign $x$ to the projection line 'Proj'.

So introducing nothing but the projection process yields,
by means of very basic projective geometry, obviously some
features known from physics usually attributed to other
contexts and pictures.\\

Nevertheless, we have further possibilities to extend this
simple geometrical setup.

\begin{figure}[h]
\includegraphics[scale=0.55]{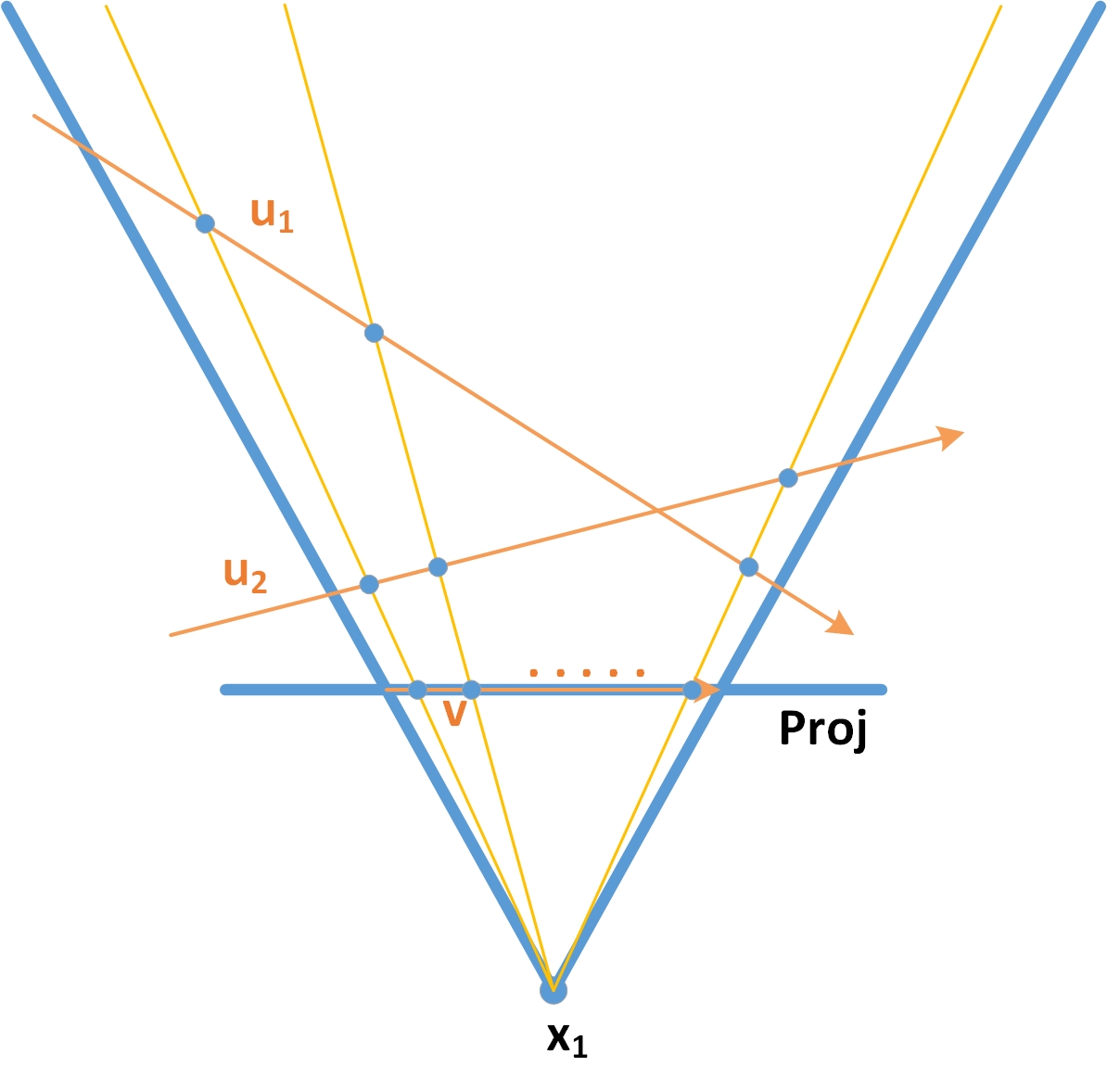}
\caption{Observer and non-parallel linear velocities.}
\label{fig:app2}
\end{figure}

If as a next enhancement, we allow for non-parallel (but 
still linear) motion in the plane (see Figure~\ref{fig:app2}),
the description still holds, however, we have to consider 
this change with respect to velocities.
Not only does the distance of the moving point, emerging 
between two points of our measurement on 'Proj', increase
at larger distance from the observer/the projection line,
it also depends on the angle between the line 'Proj' and
the direction of the linear velocity. This, however, has
to be 'absorbed' in an accelerated/decelerated motion of
the real physical motion of the point if we require the
projection to maintain its uniform velocity. In other words,
non-parallel, but linear motion decelerates as the point 
approximates the projection line/the observer, and accelerates
otherwise.

So there are two lessons to learn by this enhancement:
\begin{itemize}
\item[-] For the 'real'/ physical motion, we have to give
up uniform/constant motion in order to maintain the old
picture of the observation on the projection 'Proj', so
that for the (virtual) point moving on the line 'Proj'
the distance between two points of measurement (i.e.~the
intersections of the motion and the projection lines)
is the same at same time intervals. This enhances our
(planar) velocity classes by accelerated motions.
\item[-] From the geometrical viewpoint of Euclidean geometry,
it is now obvious, that we have to take care of intersections
of possible lines $u_{1}$ and $u_{2}$, i.e.~even in the
Euclidean description, we have to consider a second pencil
with center at the intersection of $u_{1}$ and $u_{2}$.
This is no real problem, but it shows that we have to 
switch our simple Euclidean description to a geometrically
more suitable description in terms of line pencils and 
their (projective) properties. We can then include the
case of Figure~{fig:app1} by assuming the center of the
pencil in Figure~{fig:app1} at $\infty$. In terms of 
coordinates, to achieve a unified description already 
in this simple setup, we have to give up Euclidean
coordinates and switch to projective geometry.
\end{itemize}

Having reached this stage, there are two further possibilities
to enhance our original scenario:
\begin{enumerate}
\item From the viewpoint of projective geometry, having 
two line pencils at hand, it is obvious to take possible 
projectively generated, higher order objects as well as
general projective (planar) theorems on points and line 
intersections into account. As such, we can switch to conic
sections, e.g.~to non-linear motion like in Figure~\ref{fig:app3}
for a quadratic parabola.
\begin{figure}[h]
\includegraphics[scale=0.55]{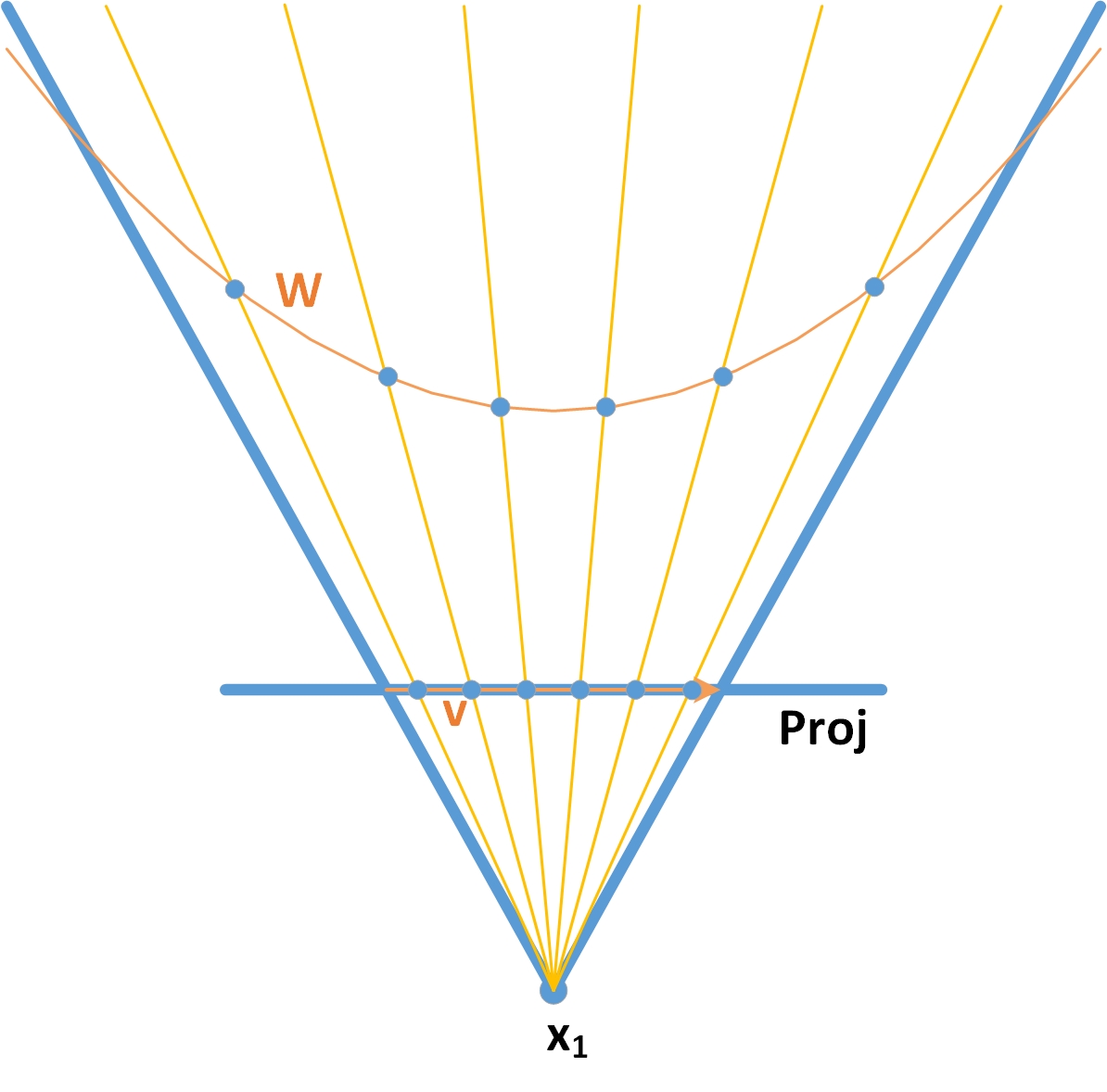}
\caption{Observer and non-linear velocities.}
\label{fig:app3}
\end{figure}
The situation is similar to Figure~\ref{fig:app2}, and we
can discuss the mapping dependent on the various parameters
of the parabola, or more generally, of second order planar
curves. However, in addition to the parameters of the curve
and the projection, we may also ask for the time deltas on 
the line 'Proj'. So also the quality of the measurement 
enters and influences the possibilities to determine the
description of motion, i.e.~for smaller $\Delta t$, we 
expect better possibilities to distinguish the various 
types of original motion.

For uniform rotation, we'll obtain trigonometric functions
(as we'll discuss in the examples of next appendix), so 
as long as the trigonometric function on 'Proj' is close
to the linearized version of the projection (within the 
quality limits of the measurement!), we find valid 
descriptions and we cannot distinguish (like e.g.~in the
case of a large radius of the rotation). So here, we
have obviously entered a regime where care has to be 
taken with respect to physical claims, although the 
mathematical treatment is well-known e.g.~in the context
of contractions and Lie algebras (see \cite{gilmore:1974},
ch.~10).

\item Last not least, we may abolish the constraint on
the observer to stay at $x_{1}$. Now, if the observer 
is allowed 'to enter' the world behind the projection
line 'Proj', he will be able to gain more information
(mostly due to his own motion), and we can apply the
considerations given in section~\ref{sec:sgr}.

Moreover, from our reasoning above, it is obvious that
the observation (if not obviously related to linear
motion), due to the projective generation of the points
of the time series, is related to planar pencils, and
thus to conic sections as projective generation of two
or more planar pencils with respect to their relative
location. In other words, we expect to see special
functions as solutions of the equations of motions 
related to such planar problems when including the 
observer.
\end{enumerate}
We have discussed in section~\ref{sec:sgr} the case 
of 'real' and 'virtual' photons already. Like above,
it is necessary to gain the freedom of moving the center
of the pencils freely by means of projective geometry,
even to an 'absolute plane', or $\infty$, respectively.

\begin{figure}[h]
\begin{tabular}{cc}
\includegraphics[scale=0.45]{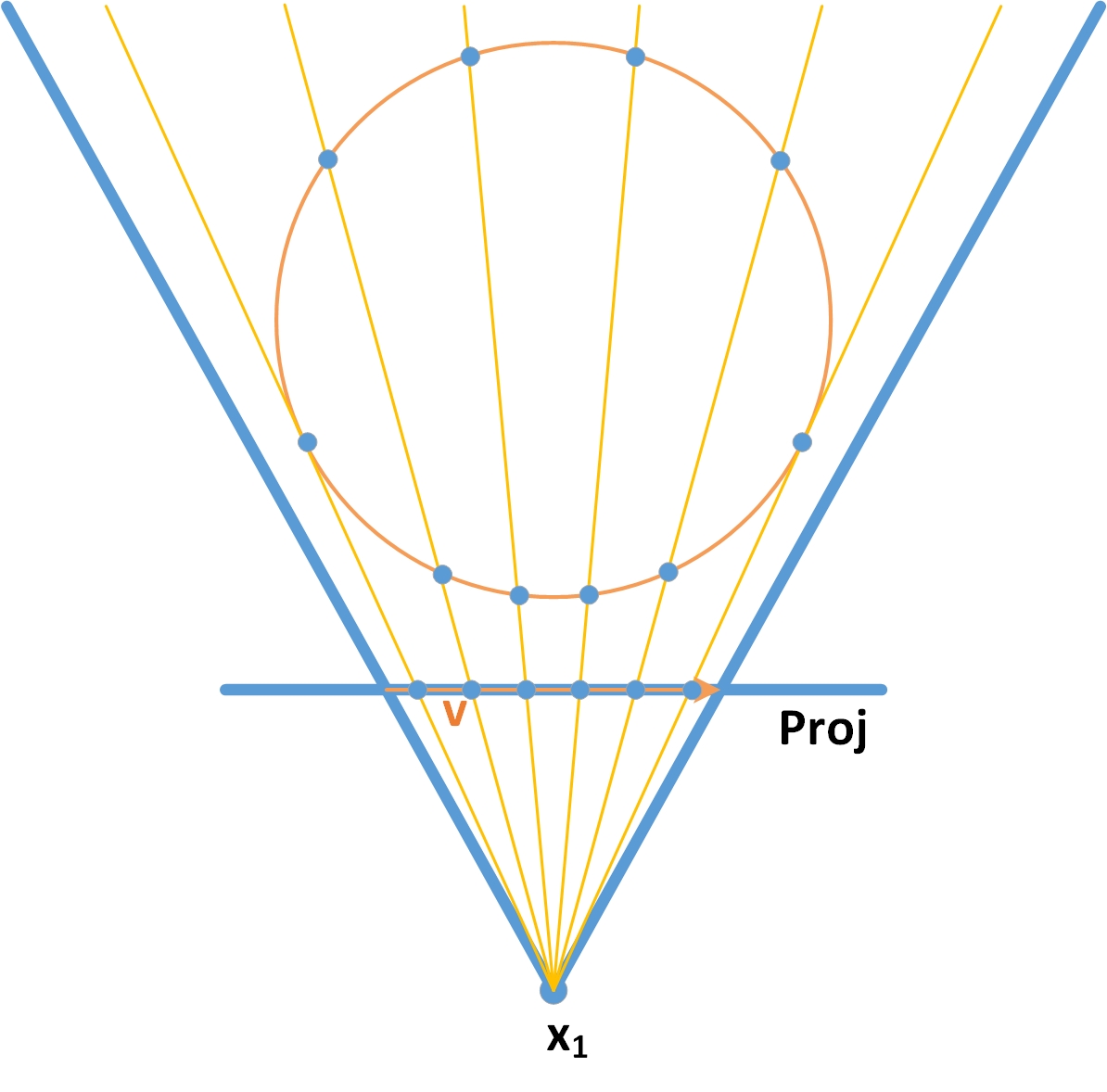}
&
\includegraphics[scale=0.45]{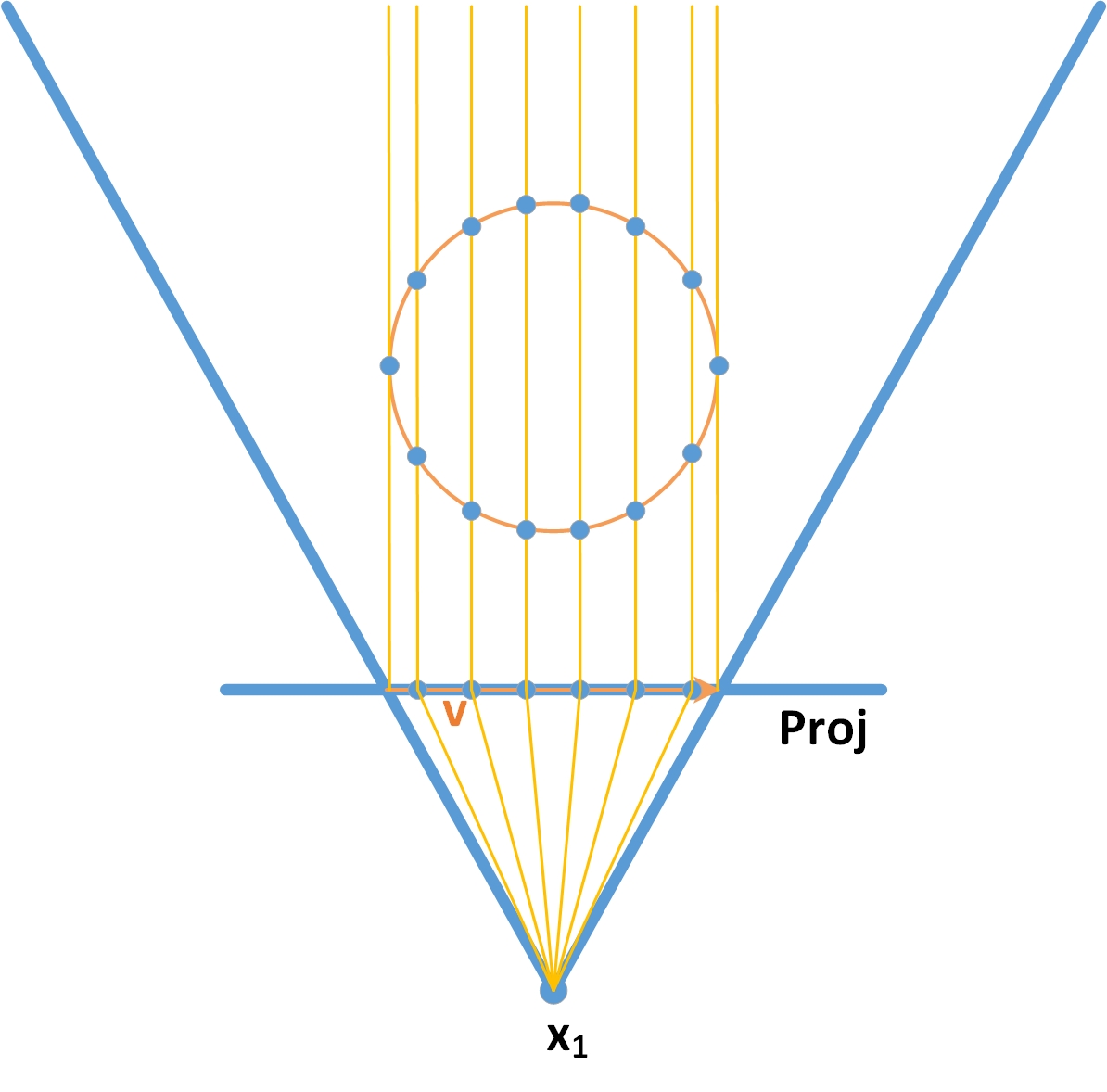}
\end{tabular}
\caption{Direct observation (left) and parallel projection
to line and observation (right).}
\label{fig:springs}
\end{figure}

Now, if instead of the parabola in Figure~\ref{fig:app3}, 
we project a circular motion Figure~\ref{fig:springs} to 
'Proj', we are in the same situation. The information 
deficit due to the projective setup causes the original
circular motion based on two variables $x$ and $y$ to 
be mapped onto one variable on the line 'Proj', only. 

Formally, this can be easily achieved by recalling the
relation of Euclidean and polar coordinates. Then, simply
extracting the $x$-part from the circular motion, we obtain
trigonometric functions to describe the power series of 
the angle $\varphi$ correctly. Dependent on the initial
point of motion/ob\-ser\-va\-tion, we thus obtain solutions 
$~\sin\varphi$ or $~\cos\varphi$. However, analytically 
we have suppressed the fact that with respect to the 
right picture, the observer is no longer located in 
$x_{1}$, but has moved to $y=-\infty$ in order to apply
the parallel projection and the related coordinate selection.
So as above in the case of the parabola, the ability of
the observer to distinguish linear from non-linear motion
depends not only on parameters of the observed objects
(e.g.~in the case of a very large radius) but also on 
the experimental setup and the quality of the observation
process.

The observation improves, and it's much easier to identify
the original process, if we move the observer from $\infty$
towards the projection line 'Proj'. Now, the lines of the 
'observer pencil' loose their parallel character more and
more, the closer the observer approaches the projection 
line, or the center of the circle with uniform motion, 
respectively. So with uniform angular motion, the circular
motion of the point to the one side (or direction) differs
more and more from the motion to the other side/direction.
Once more, the 'distortion effect' depends on the distance
of the orbit to the projection line 'Proj'.

Nevertheless, both processes should be treated consistently
in terms of line pencils and their related geometry. This 
can be seen even better if -- as above -- we move the center
of the 'observer line pencil' beyond the projection line,
or even into the orbit(s) of the motion, i.e.~close to the
center of the circular motion.\\

Last not least, we may mention the case of the circular 
motion (or the linearization as mathematical pendulum)
as physical process, and a 1-dim oscillator (or the spring,
respectively) on 'Proj'.

So as discussed already above, the information difference
between the circular motion and the projection onto the 
line 'Proj' is apparent if we consider the case of parallel
projection, i.e.~we position the observer $x_{1}$ at the
position $y\longrightarrow\infty$ in terms of Euclidean 
coordinates (or at the 'absolute line' $x_{0}=0$ of the 
plane in homogeneous coordinates).

The two related 'physical' scenarios can be summarized by
the circular planar motion with constant angular velocity
on the one hand, and the picture of a 1-dim oscillation 
(or spring) on the other hand in order to describe the 
motion of the projected point. Formally, within this 
Euclidean setup, we may just neglect the $y$-coordinate
of the circular motion by appropriately chosen origin 
in the center of the circular motion. This may be also
presented in the 'linearized' versions of the pendulum
and the spring. In all cases, however, we have to 
re-identify and rename the constants of both pictures
-- the full process and the projected oscillation -- 
with respect to their respective physical meaning.

So the parameters of the rotating point/the pendulum 
and the spring reflect in a mapping of length and 
angular velocity onto effective parameters 'mass' and
'spring constant' in the 'Proj' description, and we 
may use the dynamical picture of the rotating point 
to 'enhance' the picture of the spring. Of course, 
both physical processes exist, and nobody would 
describe springs and Hooke's law by a formally 
equivalent description in terms of a 2-dim rotation 
without need.

\end{document}